\documentclass[11pt]{article}

\usepackage[margin=1in]{geometry}
\usepackage{lmodern}
\usepackage[T1]{fontenc}
\usepackage[utf8]{inputenc}
\usepackage{booktabs}
\usepackage{amsmath}
\usepackage{graphicx}
\usepackage{xcolor}
\usepackage{colortbl}
\usepackage{multirow}
\usepackage{caption}
\usepackage{subcaption}
\usepackage{url}
\usepackage[hidelinks]{hyperref}
\usepackage{natbib}

\usepackage{placeins}

\usepackage{tikz}          
\usepackage{float}

\newcommand{\sys}{\textsc{tablet-2}}
\newcommand{\pct}[1]{#1\%}
\newcommand{\hi}[1]{\cellcolor{black!12}#1}
\newcommand{\zero}[1]{\cellcolor{red!8}#1}

\newcommand{\papertitle}{Wontopos Tablet 2:\\
Measuring Multilingual and Multimodal Memory Retrieval\\
Without Lexical Matching}

\title{\papertitle}

\author{Sunwoo Kim\\
Wontopos L.L.C.\\
\texttt{official@wontopos.com}}

\date{\today}

\newcommand{\coverpage}{%
\begin{titlepage}
\thispagestyle{empty}
\begin{tikzpicture}[remember picture, overlay]
  \node[rotate=34, text=black!7, font=\fontsize{104}{104}\selectfont\bfseries]
        at ([xshift=1.1cm, yshift=1.6cm]current page.center) {WONTOPOS};
\end{tikzpicture}\par
\vspace*{0.29\textheight}
\noindent
{\fontsize{30}{37}\selectfont Benchmark Report:\\[2pt] Tablet 2\par}
\vspace{1.6em}
\noindent
{\fontsize{13.5}{18}\selectfont\itshape\color{black!70}\papertitle\par}
\vspace{1.4em}
\noindent
{\fontsize{11}{15}\selectfont Sunwoo Kim \quad\textbullet\quad Wontopos L.L.C.\par}
\vfill
\noindent
{\fontsize{10}{13}\selectfont\today\par}
\vspace{0.7em}
\noindent\rule{0.78\textwidth}{0.7pt}\\[3pt]
{\fontsize{8}{10}\selectfont\bfseries wontopos.com}
\end{titlepage}}

\begin{document}

\coverpage

\setcounter{page}{2}

{\setcounter{tocdepth}{2}
 \tableofcontents}
\clearpage

\maketitle

\begin{abstract}
We report measurements of \sys{}, a production long-term memory engine for
language models, along two axes: the text benchmarks the field already uses,
two of them, and cross-lingual retrieval of photographs that carry no text at
all. The engine's
retrieval path contains no lexical matching, no keyword scoring, and no language
model of its own.

On LongMemEval-S (500 questions) \sys{} scores \pct{95.7}, with a 95\%
question-sampling interval of $[93.4, 97.1]$, at a median engine latency of
393\,ms per search. On BEAM-1M (700 questions over a 74{,}630-turn corpus stored
as 2.21M memories) it scores \pct{67.5}, interval $[64.8, 70.2]$.

We quote sampling intervals and not the run-to-run spread. Repeating either
benchmark varies the score by a few tenths of a point, and quoting that as
$\pm$ would state a precision the question set cannot support: the intervals
above are an order of magnitude wider. This matters for the comparison below,
where the whole spread between the top published system and ours is 7.5 points.

The rest of the text half of this paper is about how little those two numbers
mean on their own. Holding the engine, corpus, retrieval settings and judge
fixed and changing only the reader model moves LongMemEval-S by 2.0 points.
Changing only the re-ask budget moves LongMemEval-S by 1.2 points and BEAM-1M by
8.9. Neither setting is stated in the reports we compare ourselves against, and
the second is larger than most of the gaps in that comparison. We therefore
present the cross-system table as a placement and not a ranking, and give a
same-reader comparison against our Scroll-tier engine
($92.3 \rightarrow 93.7$) as the one controlled number we have.

For the multimodal axis we run two controls. Against BM25, configured as
strongly as we could make it, \sys{} reaches \pct{95.2} mean recall@5 across 70
store and query language cells on a synthetic corpus where BM25 reaches
\pct{19.0} and is exactly zero in 54 of them. On photographs stored with no
caption there is no document for a lexical method to score at all.

Because no serious competitor retrieves images lexically, we also run open
baselines on 300 Crossmodal-3600 photographs with human-written captions in 14
languages. They show that being dense does not confer language independence: an
English-trained text tower scores \pct{91.0} on English and \pct{4.7} on
Russian from identical image vectors, and a multilingual tower is level across
the languages it was trained on and collapses on Telugu (\pct{5.0}) and Swahili
(\pct{6.0}). \sys{} averages \pct{91.4} with a standard deviation across
languages of 14.0 against 27.5 and 27.7 for the baselines.

Three findings work against us and we report them at the same weight.
Retrieval degrades sharply in low-resource languages, at \pct{53.0} recall@5 for
Swahili and \pct{64.0} for Telugu, our worst numbers anywhere, though the
strongest open baseline we could run reaches \pct{6.0} and \pct{5.0} on the same
two. Attaching English captions to photographs \emph{lowers} retrieval for the
other thirteen languages by 11.4 points on average, a defect in a shipping
product and not an artifact of the benchmark: real stores hold exactly the
mixture that produces it. And one setting omitted on the way into one stage of
our own retrieval cost 37 points of Korean top-1 accuracy on short queries while
leaving nine other languages untouched. We had recorded the opposite conclusion,
that the limit was inherent to that stage, and held it for two weeks. Separating
those two cases took the same procedure both times: measure the stage on its
own, with everything around it removed, and see which number you reproduce.
\end{abstract}

\begin{table}[h]
\centering
\caption{Headline measurements. Both benchmarks were run against the same engine
build. They use different judges because each benchmark ships its own default and
we did not change either. The two scores are not comparable to each other; they
measure different corpora under different scoring rules. Delivered-token figures
are given only where we counted them with a tokeniser (\S\ref{sec:tokens}),
which on these benchmarks means BEAM-1M.}
\label{tab:headline}
\small
\begin{tabular}{llrl}
\toprule
Benchmark & Configuration & Score (\%), 95\% interval & Notes \\
\midrule
\multirow{4}{*}{LongMemEval-S}
 & Opus 5 max, re-ask 3        & \textbf{95.7} \,[93.4, 97.1] & 3 runs, 500/500 graded \\
 & GPT-5.6-sol ultra, re-ask 3 & 93.7 \,[91.1, 95.4] & 3 runs, same engine \\
 & Opus 5 medium, re-ask 3     & 95.0 & 1 run \\
 & Opus 5 medium, re-ask off   & 93.8 & 1 run, 10.5 memories read \\
\midrule
\multirow{3}{*}{BEAM-1M}
 & Opus 5 max, re-ask 3 & \textbf{67.5} \,[64.8, 70.2] & 5 runs, 3500/3500 answered \\
 & re-ask off           & 58.6 & 2 questions errored \\
 & expansion on         & 70.3 & preliminary, 681 questions, \S\ref{sec:expansion} \\
\midrule
\multicolumn{2}{l}{Crossmodal-3600, no captions, 14 languages}
                        & 91.4 & recall@5; BM25 is 0.0 \\
\bottomrule
\end{tabular}
\end{table}

\noindent
One row of Table~\ref{tab:headline} carries most of what we can say about
retrieval on its own. With re-ask switched
off entirely and the reader dropped to medium effort, the engine's first and only
retrieval scores 93.8, which edges the stronger reader running three retrieval
passes (93.7) while delivering 30\% fewer memories to it. That is the closest
thing in this paper to a measurement of retrieval quality with the reader held
still.

\section{Introduction}

A memory system for a language model has one job: given a situation, return the
things worth remembering about it. Everything downstream, meaning how the model reasons
over what it got and how it phrases an answer, is a different job, done by a
different component. That separation is easy to state and easy to lose, because
the usual way to score a memory system is to feed its output to a language model
and grade the language model's answer. What gets measured is then the pair, and
the reader can absorb a surprising amount of retrieval error.

A memory system is worth building at all only because a long context window is
not a substitute for one, and that is by now measured rather than assumed:
models use the middle of a long context least reliably \citep{liu2024lost}, and
the length they handle in practice falls short of the length they advertise
\citep{hsieh2024ruler}.

This paper reports what we can measure about \sys{}, a memory engine in
production use, along two axes: the text benchmarks that the field already uses,
and cross-lingual retrieval of photographs.

The first is the ordinary one: a benchmark, an LLM reader, an LLM judge. We use
two, LongMemEval-S at 500 questions and BEAM-1M at 700 over a corpus three
orders of magnitude larger, and report $95.7\%$ and $67.5\%$. Neither number is
worth much alone, so we also report what happens when we change the reader
model, the re-ask budget, and the context expansion policy. Those choices move
the score by more than parts of the gap between published systems, and they are
the choices other reports leave unstated.

The second axis is the one we think is under-measured. \sys{}'s retrieval path
contains no BM25, no keyword or token-overlap scoring, and no language model.
This is a design constraint, not an accident: the engine is meant to work the
same way in every national language, and lexical scoring does not. The claim is
easy to make and rarely tested, so we test it in the place where it is hardest
to fake, which is photographs stored with no caption. A photograph without a caption has
no text to match. A system that retrieves it from a text query has done something term matching
cannot do at all. We measure that on a controlled synthetic corpus in ten languages
and on 300 real photographs from Crossmodal-3600 \citep{thapliyal2022xm3600} in
fourteen, against two kinds of baseline: a BM25 one that we deliberately
configured to be as strong as we could make it, and open systems built for image
retrieval, because the interesting question is not whether term matching fails
but whether anything that avoids term matching would have done as well.

\paragraph{On comparing scores.}
Section~\ref{sec:beam} places our BEAM-1M number beside the numbers other
systems have published. We want to be explicit about what that table is and is
not. The reader model, the reader prompt, the judge model, and the retrieval
budget all move the score substantially. We measure an 8.9-point swing from the
re-ask budget alone, in Section~\ref{sec:prompt}, and those settings are
usually not reported. Every non-\sys{} number in that table is self-reported by
its authors, as is ours. We therefore treat the table as a rough placement and not a ranking. We include the table here because a paper that reports
a score with nothing beside it is not much use to a reader either.

\paragraph{Contributions.}
\begin{enumerate}
\item Measurements of a production memory engine on two text benchmarks, five
      runs and three runs respectively, with the configuration, delivered-token
      cost and latency stated, and with ablations over the reader model, the
      re-ask budget and the context expansion policy.
\item A delivered-context measurement that corrects our own earlier figures. It is counted with a tokeniser, not estimated from character counts, and checked against the character totals of the scoring runs.
\item Two controls for multilingual multimodal retrieval, on both a controlled
      corpus and real photographs: a lexical one, including the condition where
      lexical retrieval is structurally undefined, and open dense baselines
      showing that being dense does not by itself confer language independence.
\item Two negative results reported at the same weight as the rest:
      low-resource language degradation, and captions harming cross-lingual
      photo retrieval, which is a live defect rather than a benchmark
      artifact.
\end{enumerate}

\paragraph{What we do not describe.}
\sys{} is closed source, so this paper reports what it does and not how it is
built. We state what a caller can set, since a measurement cannot be read
without it, and we report what came back. The retrieval architecture, the
ranking function and the storage layout are not described.

We want to be exact about what that costs a reader, because the two things are
often confused. It costs the ability to rebuild the system from this paper. It
does not cost the ability to check the numbers: every one of them was produced
through the ordinary request interface, and the harness and the per-question
records of every run are released (Section~\ref{sec:repro}). Both benchmarks are
public, so the questions they are keyed to are obtainable from their authors. The lexical control needs no access to us at all, being
forty lines with no dependencies.

One practical limit on repeating these measurements is ours and worth stating
plainly. The runs here were made against \sys{} through the ordinary API under a
preview grant, before it was generally available. \sys{} is on general release
from 25 August 2026 and the harness is published on 29 August 2026, so from the
later of those a reader can run it against the model this paper measures; against
its predecessors, at any time after it is published. The harness and the protocol
are the same either way.

\FloatBarrier
\section{System under measurement}
\label{sec:system}

\sys{} is a store-and-retrieve engine. It accepts memories and returns memories;
it does not reason over them. Interpretation is the caller's job, or the job of
a separate component. This matters for reading the numbers below: on the text
benchmarks, a language model reads what the engine returned and writes an
answer, and the score is a property of the pair. On the multimodal measurements no
language model runs at all, because the task is to return the right photograph, and a
reader would only obscure what is being measured.

\subsection{What it does with a photograph}

A memory may carry a photograph, text, or both. \sys{} indexes the two so that
a query of either kind can reach either kind, and the property the measurements
in Section~\ref{sec:multimodal} test is the awkward direction: a photograph
stored with no caption, no title, no alt text and no surrounding message is
still returned for a text query that describes it. There is no text on that record for a query to match. That absence is what the comparison in that section rests on.

How it is done is not described here. \sys{} is a commercial product and its
retrieval is the part we sell, so this paper reports its behaviour and not its
construction.

One thing about its shape does have to be stated, because
Section~\ref{sec:attrib} turns on it. The retrieval path has more than one
stage, and the stage that decides how well a photograph matches a description is
separable from everything the engine does around it. That is all the structure
this paper uses: a weak result can sit in that stage or in the handling around
it, and the two are told apart by measuring the stage on its own.

That limit falls on the implementation and not on the measurements. Every number
below was produced through the ordinary request interface, the same one a
customer calls, so all of them can be taken again by anyone with access to the
model (\S\ref{sec:repro}). The
benchmarks are public datasets, the harness that ran them is released, and the
per-question records of our own runs are published with it
(\S\ref{sec:repro}). A reader who disagrees with a figure here can go and get
their own.

\subsection{What the caller can set}

Three request-level settings appear in the ablations below.

\begin{description}
\item[Re-ask (\texttt{verify}).] An integer from 0 to 3 that sets how many
  additional retrieval passes one request may make. Described in full in
  \S\ref{sec:reask}, because it is the setting that moves our scores most and
  the one least visible in other systems' reports.
\item[Context expansion.] Neighbouring material around a hit can be attached to
  it. Cheap in code, expensive in delivered tokens.
\item[Image budget.] The number of photographs a single response may carry. The
  default is 1 and the maximum is 5. Every recall@5 figure in this paper is
  therefore measured against a ceiling of five: ``not in the top five'' is the
  floor of what these benchmarks can observe.
\end{description}

What the engine does with a request once it has it is not described here.
Everything above is a value a caller sends, and every measurement below was
produced by sending them through the ordinary request interface.

\subsection{Re-ask}
\label{sec:reask}

Re-ask is the setting that moves our numbers most, so it is worth stating
exactly what it does and does not do. The description below is the caller-facing
contract in full. It does not describe how the engine chooses what to return.

\paragraph{The problem it addresses.}
A question carrying little context of its own, of the form ``what did I eat'',
is not reliably answered by one retrieval. A person in that situation recalls
the answer once given a hint. The equivalent here is to search again, told what
has already been seen, so the second pass can return something the first did
not.

\paragraph{Relation to iterative retrieval.}
Searching more than once for a single question is an old idea. Multi-hop dense
retrieval answers a complex question with a sequence of retrievals rather than
one \citep{xiong2021mdr}, self-ask has the model explicitly ask and answer
follow-up questions before the original one, with a search engine plugged in to
answer them \citep{press2023compositionality}, and IRCoT interleaves retrieval
with chain-of-thought reasoning steps \citep{trivedi2023ircot}. What those share
is that a language model composes the next query. Re-ask does not, and the
distinction is narrower than it first looks, so it is worth stating precisely. A
caller may put a model in its own loop to decide \emph{whether} to spend another
pass, and in one of our two campaigns we did exactly that. What no model does at
any point is change \emph{what} the next pass looks for: a later pass is the same
query with the memories already seen excluded, and the engine itself never calls
a model. The trade is explicit. We give up query reformulation, and in exchange
retrieval stays deterministic given the request, can be priced before it runs,
and needs no model credentials of its own. It is also why we report the question
type where re-ask loses points rather than only the average.

\paragraph{What the caller sets.}
One integer, \texttt{verify}, between 0 and 3. It is the number of
\emph{additional} passes permitted, so \texttt{verify:2} allows at most three
retrievals in total. The default is 0, so callers who do not set it get exactly
one retrieval and are billed for one. A value outside the range, or of the wrong type, is
rejected with a 400 before the engine is called, and not quietly clamped: a caller who sends 5 and is silently given 3 has no way to learn that.

\paragraph{What happens between passes.}
Each subsequent pass carries the pass number and the identifiers of the memories
already returned, and the engine excludes them. Results accumulate across passes
and a memory returned twice appears once. The engine holds no state between passes. The state travels in the request, so any pass can be retried or served by a different instance.

\paragraph{The engine does not call a language model.}
Not on the first pass and not on any later one. This is the property the whole
design protects. A memory system that called an LLM to decide whether to search
again would need the caller's model credentials, would add that model's latency
to every retrieval, and would make the retrieval path non-deterministic.

\paragraph{\texttt{verify} is a ceiling, not a schedule, and this matters for
reading our two campaigns.}
The parameter caps how many additional passes a request may make. It does not
say that they will be made. The engine stops early when a pass returns nothing
new, and above that floor the decision to spend another pass belongs to whatever
loop the caller runs. What sits in that loop is the caller's business, and ours
differed between the two benchmarks, which is the single most important thing to
know before comparing the two re-ask ablations:

\begin{description}
\item[BEAM-1M.] The budget is set once and applied to every question. No
  component outside the engine decides anything per question, so a run
  reproduces without a reader at all, which is what \S\ref{sec:tokens} relies on.
\item[LongMemEval-S.] The reader was in the loop. It read what came back and
  asked again when it judged the evidence insufficient, up to the same ceiling.
  The budget is therefore an upper bound that different readers spend
  differently, which is what Table~\ref{tab:hops} measures and what makes the
  reader comparison in \S\ref{sec:lme} a comparison of two things at once.
\end{description}

\noindent
The engine is identical in both and reasons in neither. The difference is
entirely above it. We separate the two here because the same integer buys 8.9
points on one benchmark and 1.2 on the other, and part of that gap is this.

\paragraph{It stops early, and says so.}
If a pass returns nothing new, the remaining passes are not performed and not
billed. The response reports \texttt{verify\_used}, the number of additional
passes actually made, which can be lower than the number requested. A caller who asks for 3 and is told 1 learns that the corpus had nothing further to give, without having to infer it from a bill.

\paragraph{Failure is partial, not total.}
If a later pass fails, the results already gathered are returned rather than the
whole request failing. A re-ask that breaks should cost a caller some recall, not
their answer.

\paragraph{Availability is checked, not assumed.}
Not every engine version implements the protocol, and one that does not will
accept the extra fields, ignore them, and answer 200. The loop would read that as
``nothing new'' and stop, so the caller would have paid for a re-ask that never
happened and been told nothing. Asking for re-ask on such a model is therefore
refused outright, before any engine call, so that the refusal precedes the
charge.

\paragraph{What it costs and what it buys.}
On BEAM-1M, \texttt{verify:3} is worth 8.9 points at 3.2 times the delivered
context (\S\ref{sec:prompt}). On LongMemEval-S it is worth 1.2 points at roughly
twice the memories (\S\ref{sec:lme}). The difference between those two figures is
the useful part: re-ask pays in proportion to how often one retrieval misses, so
it earns its cost on a 2.2M-memory corpus and much less on a small one. It is
also not uniformly positive. On LongMemEval-S's \texttt{single-session-user}
questions it costs 4.2 points, because asking again when the answer was already
present adds material that competes with it.

\paragraph{Interface note.}
The client libraries forward request options they do not recognise instead of
rejecting them, so \texttt{verify} is set through the ordinary search options
and has been callable since well before this measurement. It is not yet a named
argument with documentation of its own. A reader who wants the current parameter
reference should consult the API documentation at \url{https://wontopos.com};
this paper fixes it at the state we measured.

\FloatBarrier
\section{Benchmark 1: LongMemEval-S}
\label{sec:lme}

\subsection{Setup}

LongMemEval-S \citep{wu2024longmemeval} poses 500 questions across six
categories over multi-session dialogue histories. Scoring is binary per question, decided by a judge model
that receives the question, the ground truth, and the answer, and returns only
yes or no. The judge is \texttt{gpt-4o} at temperature 0, which is this
benchmark's default; we did not change it. Answer generation and judging run as
separate processes, so no model grades its own output. Grading free-form answers
with a model, rather than by string match, is the standard both benchmarks
assume \citep{zheng2023judge}; \S\ref{sec:limits} states what that costs us.

Table~\ref{tab:gen} lists the engines this section compares.
Retrieval settings were fixed across every run reported in this section:
20 results per search, context expansion on, structural and duplicate filtering
on, archival lane on. The engine binary was the production build, copied to the benchmark machine and not rebuilt there.

\textbf{Note that this differs from our BEAM-1M configuration}, where the
headline number is measured with expansion off (\S\ref{sec:beam}). The two
campaigns were run months apart and we have not re-run either under the other's
settings. A reader comparing the two scores should treat them as measurements of different systems-under-test as well as different corpora. That is a further reason the two numbers do not belong side by side.

\subsection{Result, and what moves it}

\begin{table}[t]
\centering
\caption{LongMemEval-S. Every configuration we ran. The two main configurations
differ only in the reader; the engine, corpus, retrieval settings and judge are
identical. Brackets are Wilson 95\% intervals \citep{wilson1927interval} at one
run's worth of questions; scoring here is one verdict per question, so a binomial
interval is the right one, and the three runs are not pooled because they are the
same questions asked again. \emph{Graded} is the denominator: the judge returned
no verdict on one question in each GPT-5.6-sol run, and an ungraded question
leaves the denominator rather than counting as wrong (\S\ref{sec:lmeprotocol}).}
\label{tab:lme}
\begin{tabular}{llrrrl}
\toprule
Reader & Effort, re-ask & Runs & Graded & Score (\%) & Per-run \\
\midrule
Claude Opus 5 & max, 3    & 3 & 500 & \textbf{95.7} \,[93.4, 97.1] & 95.2 / 96.0 / 96.0 \\
GPT-5.6-sol   & ultra, 3  & 3 & 499 & 93.7 \,[91.1, 95.4] & 93.2 / 94.0 / 93.8 \\
Claude Opus 5 & medium, 3 & 1 & 500 & 95.0 \,[92.7, 96.6] & \\
Claude Opus 5 & medium, 0 & 1 & 500 & 93.8 \,[91.3, 95.6] & \\
\bottomrule
\end{tabular}
\end{table}

\paragraph{Why three runs here and five on BEAM-1M.}
Both three-run configurations came in above 94\% with a run-to-run spread of
0.8, so a fourth and fifth repeat of either would have narrowed an interval that
was already narrow relative to every other quantity in this section. The budget
went to varying the configuration instead: two reader models, two effort levels
within one of them, and the re-ask budget swept from off to three. That gives four points of comparison instead of one number measured more precisely, and it is what makes the reader effect below measurable at all. BEAM-1M is repeated five
times because its score is the one we place beside other systems, and there the
precision is the point.

The two three-run configurations do not overlap in their runs: the lowest Opus 5
run (95.2) is above the highest GPT-5.6-sol run (94.0), and both have a
run-to-run spread of 0.8. \textbf{Swapping the reader is worth 2.0 points} with
everything else held constant.

Their sampling intervals do overlap, and we say so: at $n=500$ a 2.0-point
difference near \pct{95} is inside what 500 questions can resolve. What makes
this measurement usable anyway is that it is paired. Both readers answered the
same 500 questions on the same retrievals, so the question-to-question variance
that dominates the interval is shared and cancels; the six runs separate cleanly.
A reader who wants the effect established on independent question sets would
need a larger benchmark than this one. This is the measurement we rely on in \S\ref{sec:beam} when
reading cross-system comparisons, and it is why we regard those comparisons as
loose.

\subsection{Isolating retrieval from the reader}

A reader swap does not only change judgement; it changes how often the reader
asks the engine for more. Table~\ref{tab:hops} shows that the stronger reader
concluded its evidence was insufficient more than twice as often, and therefore
answered from 40\% more material. Attributing the full 2.0 points to
``better reasoning over the same input'' would be wrong, because the input was
not the same.

\begin{table}[t]
\centering
\caption{Retrieval behaviour on the same 500 questions. Re-asking is the
reader's decision, so different readers consume different amounts of engine
output.}
\label{tab:hops}
\begin{tabular}{lrrrr}
\toprule
Configuration & Mean passes & Re-asked & Memories read & Characters \\
\midrule
GPT-5.6-sol ultra   & 0.65 & 210/500 (42\%) & 14.9 & 28{,}037 \\
Opus 5 max          & 1.62 & 430/500 (86\%) & 21.2 & 41{,}000 \\
Opus 5 medium       & 1.43 & 477/500 (95\%) & 19.6 & 36{,}951 \\
Opus 5 medium, off  & 0 & 0/500 & 10.5 & 19{,}175 \\
\bottomrule
\end{tabular}
\end{table}

Disabling re-ask removes that confound, because the reader then answers from
whatever the first search returned and cannot ask for more.

\begin{table}[t]
\centering
\caption{With re-ask disabled, the weaker configuration matches the stronger one
on 30\% fewer memories.}
\label{tab:hop0}
\begin{tabular}{lrr}
\toprule
& Score (\%) & Memories read \\
\midrule
Opus 5 medium, re-ask off & 93.8 & 10.5 \\
GPT-5.6-sol ultra, re-ask 3 & 93.7 & 14.9 \\
Opus 5 medium, re-ask 3 & 95.0 & 19.6 \\
\bottomrule
\end{tabular}
\end{table}

\begin{figure}[t]
\centering
\includegraphics[width=\textwidth]{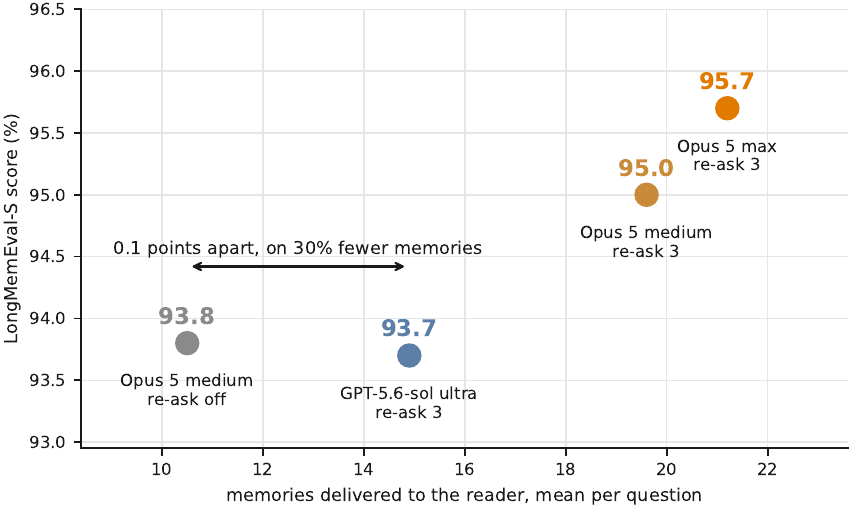}
\caption{The reader is not a constant. Engine, corpus, retrieval settings and
judge are identical across all four points; only the reader model and the
re-ask budget move. Horizontal position is what the engine handed over, so a
point that sits further right cost the reader more context to reach the same
height. Data from Tables~\ref{tab:lme} and~\ref{tab:hops}.}
\label{fig:reader}
\end{figure}

We read Table~\ref{tab:hop0} and Figure~\ref{fig:reader} as the cleanest statement available about the
engine alone. A mid-effort reader, given one retrieval and no opportunity to ask
again, scores level with a maximum-effort reader that took three passes. Re-ask
then adds 1.2 points on top (93.8 to 95.0). On BEAM-1M the same setting is worth 8.9 points (\S\ref{sec:beam}). Re-ask pays in proportion to how often the first retrieval misses.

Re-ask is not uniformly good. On \texttt{single-session-user} it costs 4.2
points (97.1 to 92.9) while helping four of the other five categories. Asking
again when the first answer was already present adds material that competes
with it.

\subsection{Same-reader comparison against our other engine tier}

Cross-generation claims are usually confounded by reader changes. Here we have
one that is not.

\begin{table}[t]
\centering
\caption{Our engines on LongMemEval-S. Tablet and Scroll are separate tiers
rather than successive generations; scroll-1.2 is the larger-context tier and
tablet-1 is \sys{}'s own predecessor. The final row is the controlled
comparison: identical reader, so the difference is the engine. Delivery figures
are omitted; the earlier generations were measured in their own campaigns and we
did not re-derive them for this paper.}
\label{tab:gen}
\begin{tabular}{llr}
\toprule
Engine & Reader & Score (\%) \\
\midrule
tablet-1   & GPT-5.5      & 85.2 ($\pm$1.1, 5 runs) \\
scroll-1.2 & GPT-5.6-sol  & 92.3 \\
\sys{}     & Opus 5 max   & \textbf{95.7} \\
\midrule
\sys{}     & GPT-5.6-sol  & 93.7 \\
\bottomrule
\end{tabular}
\end{table}

Against scroll-1.2 read by the same model, \sys{} gains 1.4 points. Earlier
comparisons in our own records had to be described as lower bounds because the
readers differed; this one does not.

The same caveat applies as above and is larger here, because these two numbers
come from separate campaigns rather than paired runs. At $n=500$ the intervals
are $[89.7, 94.4]$ and $[91.1, 95.4]$ and they overlap substantially. We report
1.4 points as what we measured with the reader held fixed, and not as an
established difference.

\subsection{Latency, and a figure we withdraw}

Over 2{,}500 searches at 20 results each the engine reported a median search
latency of 393\,ms (p95 678\,ms, mean 441\,ms) and returned 19 memories at the
median. The client observed 397\,ms for the same searches, so on this deployment
the transport contributed 4\,ms.

Variation between runs exceeded variation within them. Four of the five runs had
medians between 357 and 414\,ms and the fifth sat at 501\,ms with a 3.6\,s
maximum. We have no explanation for that run and have not excluded it.

An earlier summary of ours gave 405\,ms. It pooled a sixth file that had been
recorded while a scoring run was executing on the same machine, and which
carries a note from the script that wrote it saying its latency columns are
contaminated and should not be used. The figure above excludes it.

\paragraph{We withdraw the delivered-token figure for this benchmark.}
We previously reported 978 tokens per search here, and it is the wrong quantity
to put beside a score. It counts one search, while a question answered with
three retrieval passes receives more than one search's worth; the per-question
total was never measured. For that reason we are not restating it from the same records. The counted per-question figures in \S\ref{sec:tokens} are
for BEAM-1M.

\subsection{Protocol notes}
\label{sec:lmeprotocol}

We record the following because each one could have produced a wrong number and
two of them nearly did.

\begin{enumerate}
\item \textbf{Two runs invalidated.} During a three-run sequence the second and
  third hit a session limit on the reader API, and all 500 questions returned
  the provider's limit message. Those runs were quarantined and re-measured.
\item \textbf{The limit message was not empty}, so it passed our
  ``count the blank answers'' validity check. We changed the check to count
  non-answers, not just empty strings.
\item \textbf{The three valid Opus 5 runs span two session windows.} Successful
  responses were cached and only blocked questions were re-asked in the next
  window. Model, effort and engine were identical throughout, but the runs were
  not each executed in one continuous window.
\item \textbf{A measurement bug inflated a published figure.} Our script sorted
  latencies but not token counts, so the reported token median was the value at
  the midpoint of the unsorted list rather than the median, an overstatement of
  60\%. It was recomputed after the fix, and the corrected figure is the one we
  withdraw above for a separate reason.
\item \textbf{One run was invalid and looked fine.} A relay key mismatch meant
  the reader received zero memories for all 500 questions and answered anyway;
  the logs showed no error. Every run since checks the retrieved count before
  grading.
\item \textbf{No per-question comparison against tablet-1 exists.} We produced
  one, then withdrew it after finding the archived file we had used was not the
  configuration we believed. A real generation comparison would need tablet-1
  re-run on the same corpus, reader and re-ask budget.
\item \textbf{One question per GPT-5.6-sol run has no verdict.} Each of those
  three runs returned 499 gradable answers and one the judge did not score. An
  ungraded question leaves the denominator rather than counting as wrong, since
  scoring a grader failure as an engine failure would make an outage look like a
  result. That row is therefore 499 questions and the Opus 5 rows are 500, which
  is why Table~\ref{tab:lme} states each row's denominator. The choice is worth
  1 question in 500, or 0.2 points, and it moves in our favour.
\end{enumerate}

\FloatBarrier
\section{Benchmark 2: BEAM-1M}
\label{sec:beam}

\subsection{Setup}

BEAM-1M poses 700 questions over a corpus of 35 conversations totalling 74{,}630
turns, which the engine stored as 2{,}212{,}504 memories. Ten question types are
represented, from preference following to event ordering.

Scoring follows the benchmark's own protocol. Each question carries a rubric of
nugget items; a judge marks each item $1.0$, $0.5$, or $0.0$, and the question
score is the mean over its items. Decomposing an answer into atomic facts and
scoring each one separately, rather than grading the answer whole, is the nugget
tradition in retrieval evaluation, and recent work automates the decomposition
with a language model in much this form \citep{pradeep2025nuggets}. The one
exception is \texttt{event\_ordering}, which the benchmark scores by rank
correlation and not by rubric mean; we use its normalised Kendall $\tau_b$
\citep{kendall1945ties}, as its reference implementation does, the $b$ variant
being the one that corrects for tied ranks. This
distinction is worth flagging because scoring \texttt{event\_ordering} with the
rubric mean instead, which we did at first, inflates the overall score by
roughly four points.

\begin{table}[t]
\centering
\caption{Measurement configuration. The reader prompt is ours, not the
benchmark's; Section~\ref{sec:prompt} reports both.}
\label{tab:config}
\begin{tabular}{ll}
\toprule
Engine & \sys{}, same binary as production \\
Request & re-ask $=3$, limit $=20$, no context expansion, two lanes \\
Reader & Claude Opus 5, maximum reasoning effort \\
Judge & \texttt{gpt-4.1-mini}, temperature 0, benchmark's own judge prompt \\
Aggregation & rubric mean (630 questions) $+$ normalised Kendall $\tau_b$ (70) \\
Corpus & 35 conversations, 74{,}630 turns $\rightarrow$ 2{,}212{,}504 memories \\
\bottomrule
\end{tabular}
\end{table}

Table~\ref{tab:config} states the configuration in full. We used
\texttt{gpt-4.1-mini} as judge because it is the default in the
benchmark's own scoring code and we did not change it. Some published results
on related benchmarks use a larger judge; we note the difference and did not adjust for it, since adjusting would mean choosing a judge after seeing scores.

\subsection{Result}

\begin{table}[t]
\centering
\caption{Five runs of the same configuration. 3{,}500 questions answered, zero
missing answers. The spread across runs is 0.7 points; it describes how much
re-running the same 700 questions moves the score, and is not the uncertainty of
the score. That is given below.}
\label{tab:runs}
\begin{tabular}{lrrrrrr}
\toprule
Run & 1 & 2 & 3 & 4 & 5 & Mean \\
\midrule
Score & 67.8 & 67.7 & 67.3 & 67.1 & 67.5 & \textbf{67.5} \\
\bottomrule
\end{tabular}

\vspace{4pt}
\begin{tabular}{ll}
\toprule
Run-to-run spread ($\sigma$) & 0.22 points \\
95\% question-sampling interval & $[64.8, 70.2]$, i.e.\ $\pm2.7$ points \\
\bottomrule
\end{tabular}
\end{table}

\noindent
Table~\ref{tab:runs} gives the five runs. The two rows below the scores differ
by more than a factor of ten, and only the
second is an uncertainty. BEAM scores each question on a continuous scale rather
than as right or wrong, so the interval is a bootstrap over questions
\citep{efron1979bootstrap} at 2{,}000 resamples, and not a binomial one; the
Wilson interval used in \S\ref{sec:lme} would not apply here, because there is no
Bernoulli trial to count. Run-to-run variation is small because the
retrieval is close to deterministic; it says nothing about how much the score
would move on a different 700 questions drawn the same way.

\begin{table}[t]
\centering
\caption{By question type, mean of five runs.}
\label{tab:bytype}
\begin{tabular}{lrr}
\toprule
Question type & Mean (\%) & $\sigma$ \\
\midrule
preference\_following      & 90.4 & 1.2 \\
abstention                 & 83.4 & 2.5 \\
information\_extraction    & 78.2 & 2.1 \\
contradiction\_resolution  & 78.0 & 0.9 \\
knowledge\_update          & 75.4 & 2.2 \\
instruction\_following     & 74.7 & 2.8 \\
multi\_session\_reasoning  & 63.8 & 1.2 \\
temporal\_reasoning        & 54.8 & 2.4 \\
summarization              & 52.5 & 0.3 \\
event\_ordering            & 23.6 & 0.5 \\
\bottomrule
\end{tabular}
\end{table}

\begin{figure}[t]
\centering
\includegraphics[width=\textwidth]{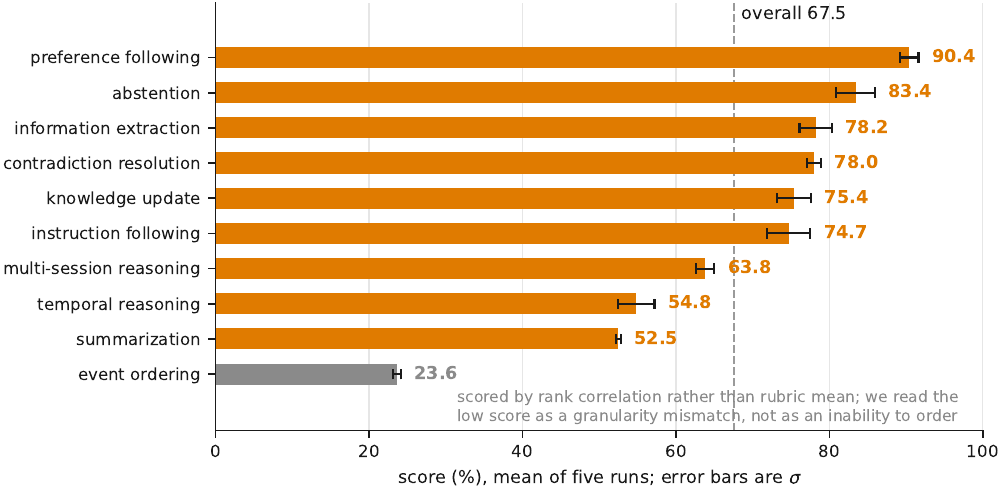}
\caption{BEAM-1M by question type, the data of Table~\ref{tab:bytype} drawn
against the overall mean. The run-to-run spread is small enough that the
ordering of types is stable; the interesting variation is between types, not
between runs. \texttt{event\_ordering} is drawn in grey because it is the one
type scored by rank correlation and not by rubric mean, which
\S\ref{sec:beam} argues makes its value not comparable to the others.}
\label{fig:bytype}
\end{figure}

\subsection{Delivered tokens}
\label{sec:tokens}

Delivered context is the cost a memory system imposes on whatever reads it, so
it belongs beside the score. The cost is not only money and latency: a reader
handed more material does not use all of it evenly, and evidence placed in the
middle of a long context is used less reliably than evidence at either end
\citep{liu2024lost}. A memory system that reaches its score by delivering more
has moved work onto the component least able to refuse it. It is also easy to get
wrong, and we did.

Our benchmark harness counts tokens only when it is given a tokeniser command,
and never estimates one, on the principle that an approximate token count is
worse than none. The five scoring runs were executed without a tokeniser
attached, so every row recorded characters and a null token count. The token
figures we published alongside those runs were produced afterwards by dividing
characters by a constant. Comparing them against a competitor's counted tokens
was therefore not a valid comparison, whichever way it came out.

We re-measured. Ten questions from each of the ten question types, 100 in total,
were run through the same harness against the same corpus and engine, with a
null reader in place of the language model and with
\texttt{tiktoken}'s \texttt{o200k\_base} attached as the tokeniser. No answer was
generated and nothing was graded; the only quantity of interest is what the
engine hands back. The re-ask budget is a request parameter and not a reader decision, so the multi-pass configurations reproduce without a reader.

The check that makes these numbers usable is the character count. If the
re-measurement assembles context the same way the scoring runs did, its
character totals must match theirs. They do: a median of 3{,}902 characters
against the 3{,}884 recorded for the same configuration across 700 questions.

\begin{table}[t]
\centering
\caption{Delivered context per query, counted rather than estimated
(\texttt{o200k\_base}, 100 questions, 10 per type, medians). The final column is
the character median recorded during the 700-question scoring runs, and is the
control: a re-measurement that assembles context differently would not land
near it.}
\label{tab:tokencorrect}
\begin{tabular}{llrrrr}
\toprule
& Configuration & Characters & Tokens & Previously published & Scoring runs \\
\midrule
A & re-ask off & 3{,}902 & 776 & 1{,}335 & 3{,}884 \\
B & re-ask 3 \quad\emph{(headline)} & 12{,}200 & 2{,}484 & 4{,}107 & 11{,}999 \\
C & expansion on & 74{,}710 & 16{,}590 & 22{,}064 & 64{,}204 \\
\bottomrule
\end{tabular}
\end{table}

Table~\ref{tab:tokencorrect} sets the counted figures beside what we had
published. The true ratio is 4.5 to 5.0 characters per token on this corpus, not the 2.91
that had been assumed, so the published figures overstated our delivery by 33 to
72\%. The correction moves in our favour. That is the direction a reader should check hardest, so the character counts are given alongside:
any reader can apply their own tokeniser to them.

Configuration C reproduces less tightly than A and B, at 74{,}710 characters
against the 64{,}204 recorded. Expansion attaches neighbouring material to each
hit, so the amount delivered depends heavily on which memories a particular
question retrieves, and a 100-question sample carries more of that variance than
a 700-question run. We report C as approximate for that reason.

\paragraph{A difference between builds.}
These re-measurements were run against the same engine binary that produced the
scores (SHA-256 prefix \texttt{cfee6b9d}). We first ran them against a newer
build of the engine and configuration C delivered 17{,}476 characters, a
quarter of what the scoring build delivers, while A and B were unchanged. We do
not yet know whether the newer build fixes an over-delivery or disables
expansion, and it is not the build these scores were measured on, so the numbers
above are from the scoring build. We flag it because it is the kind of
difference that a hash comparison catches and a re-run without one does not.

\subsection{Latency}

The engine's own latency is reported in \S\ref{sec:lme}, measured on the LongMemEval-S corpus, and is not repeated here: the figure belongs to the
engine rather than to a benchmark, and that campaign is where we have the raw
per-search records to compute it from.

What belongs here is the rest of the stack. Two further layers sit between
engine and caller in our deployment, and measured separately the backend adds
284\,ms and the preview proxy 66\,ms. Production customers pay the backend's share too, so it is not a benchmark artifact and we give it alongside the engine figure.

The BEAM harness in these runs also crossed a long network hop, a property of where we ran it and not of the system. The wall-clock times recorded per question in the released runs are therefore not engine latencies.

\subsection{Placement against published numbers}

\begin{table}[t]
\centering
\small
\caption{BEAM-1M scores at the 1M scale as published by their respective
authors, retrieved August 2026. Every number including ours is self-reported.
The reader column is the answering model where the source states it. Note that
no two rows are known to share a reader.}
\label{tab:compare}
\begin{tabular}{lrlll}
\toprule
System & Score (\%) & Tokens/query & Reader & Source \\
\midrule
Exabase M-1     & 75.0 & not stated$^{\dagger}$ & Gemini 3 Flash & own, Jul 2026 \\
Hindsight       & 73.9 & not stated & Gemini 3 Pro & own, Apr 2026 \\
\textbf{\sys{}} & \textbf{67.5} & \textbf{2{,}484} & Claude Opus 5 & this paper \\
Mem0            & 64.1 & $\approx$6{,}900 & not stated & own, 2026 \\
Honcho          & 63.1 & not stated & Gemini 3 Pro & via Hindsight \\
LIGHT           & 33.6 & n/a & not stated & via Hindsight \\
RAG baseline    & 30.7 & n/a & not stated & via Hindsight \\
\bottomrule
\end{tabular}

\vspace{2pt}
{\footnotesize $^{\dagger}$Exabase reports consuming about 20\% fewer tokens per
query than the next best system, without giving the count.}
\end{table}

Read as a ranking, \sys{} places third of the six named systems in
Table~\ref{tab:compare}. We do not think
that reading is supportable, and this paper contains the measurements that say
why.

The reader is different in every row where it is known. Exabase states plainly
that it used Gemini 3 Flash while the other leading systems used Gemini 3 Pro;
we used Claude Opus 5. On LongMemEval-S we measured a reader swap to be worth
2.0 points with everything else held fixed (\S\ref{sec:lme}), and in
\S\ref{sec:prompt} we measure the re-ask budget to be worth 8.9 points on BEAM
itself. The gap between the top published system and ours is 7.5
points, and our own question-sampling interval is $\pm2.7$ before any of that is
considered. Two settings that nobody reports are each capable of moving a system
through most of what remains.

\paragraph{One row in that table runs this argument against us.}
It would be convenient to leave the Flash detail as a fact and move on, so we
will not. Exabase's report is explicit that it reached 75.0 on the cheaper
reader tier while the other leading systems used Gemini 3 Pro, and it presents
that as the result rather than as a caveat. If the reader moves a score as much
as we measure it to, then a system reaching 75.0 without the stronger reader was
not carried there by its reader, and the 7.5 points between that row and ours
are not explained by the variable we have just spent three paragraphs pointing
at. The observation that these numbers are uncontrolled is symmetric. On this
row it points away from us, and a reader of this paper would have worked that
out whether or not we wrote it down.

What we can say is narrower than we would like, and it is not that their number
is soft. It is that the table cannot be read as a ranking in either direction,
including the directions that would suit us. That report gives one figure per
scale and states no repeat count, no interval, no judge, no prompt and no
retrieval budget; its token claim is a relative saving against an unnamed
neighbour rather than a count, so it cannot be set beside our 2{,}484. We give
five runs, a spread of 0.22, a bootstrap interval of $\pm2.7$ and the
configuration in Table~\ref{tab:config}, which makes our number checkable and
does nothing to make it larger. The measurement that would actually settle the
comparison is our engine run under their reader and their prompt, and we have
not run it. Until we do, the honest summary of Table~\ref{tab:compare} is that a
system reports a higher score than ours on a cheaper reader and we have not
measured why.

The judge is a second unreported variable. The benchmark's authors do not fully
standardise the judge or the evaluation prompts, a limitation Mem0 has noted in
print as well. We used the benchmark's default judge without changing it. That is the most defensible choice available to us and still not the same as knowing the other rows did likewise.

We include the table because a score published with nothing beside it is not
useful either. On our own product pages we publish only our own
measurement, with the reader and effort level named on each row.

\subsection{Ablation: re-ask and expansion}
\label{sec:prompt}

\begin{table}[t]
\centering
\caption{Configuration ablations. B is the headline measurement. Token figures
are counted (\S\ref{sec:tokens}).}
\label{tab:ablation}
\begin{tabular}{llrrl}
\toprule
& Configuration & Score (\%) & Tokens & \\
\midrule
A  & re-ask $=0$                        & 58.6 & 776 & 2 questions errored \\
B  & re-ask $=3$ \quad\emph{(headline)} & 67.5 & 2{,}484 & 5 runs \\
C  & re-ask $=3$ $+$ expansion          & 70.3 & 16{,}590 & preliminary, 681 questions \\
\bottomrule
\end{tabular}
\end{table}

\begin{figure}[t]
\centering
\includegraphics[width=\textwidth]{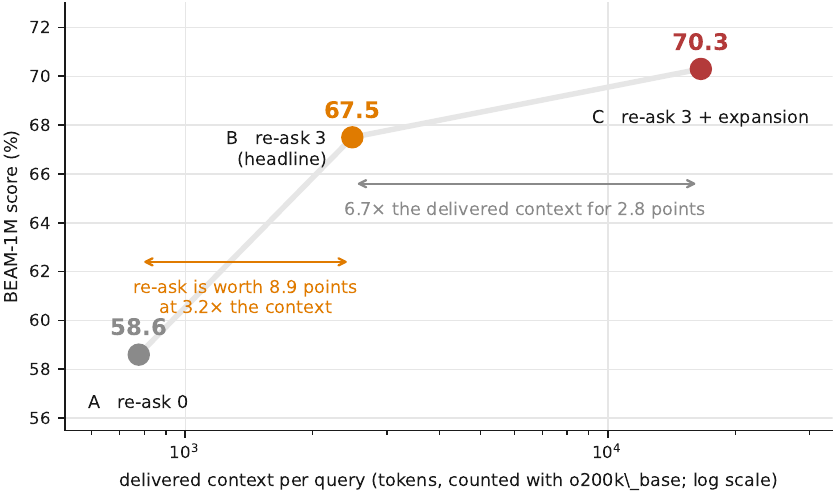}
\caption{What context buys on BEAM-1M. Score against delivered context for the
three configurations of Table~\ref{tab:ablation}, with tokens counted rather
than estimated (\S\ref{sec:tokens}). The horizontal axis is logarithmic, which
is the only way to put A and C on one plot, and is also the point: the last
2.8 points cost an order of magnitude more context than the first 8.9.}
\label{fig:tokens}
\end{figure}

\paragraph{The reader prompt is ours, not the benchmark's.}
BEAM does not fix a reader model or require its own reader prompt; the field for
the answering model in its generation script is left empty. Ours is 192
characters long and was not tuned against this benchmark: it says to answer from
the memories only, not to use outside knowledge, to say so when the memories do
not contain the answer, and to be concise. It is reproduced verbatim in
Appendix~\ref{app:prompts}. The benchmark ships a longer reader prompt of its
own for RAG baselines, which we did not use and do not reproduce here; it is in
its public repository.
Anyone reproducing these numbers with a different reader prompt should expect a
different score, and we say so here rather than leaving it to be discovered.

Figure~\ref{fig:tokens} plots the three configurations against what they cost.

\paragraph{Re-ask is worth 8.9 points.}
Turning it off costs $58.6$ against $67.5$. The same setting is worth 1.2 points
on LongMemEval-S (\S\ref{sec:lme}). The gap between those two figures is itself
informative: re-ask pays in proportion to how often the first retrieval misses,
and on a 2.2M-memory corpus it misses far more often than on a 500-question one.

\paragraph{Expansion works, and is not usable as it stands.}
\label{sec:expansion}
Configuration C turns on context expansion, which attaches neighbouring material
to each retrieved memory, along with duplicate filtering and the archival lane.
It scores highest of the three at $70.3\%$. We do not present it as our result,
for three reasons.

The delivery cost comes first. C hands the reader 16{,}590 tokens per query against
2{,}484 for B: $6.7\times$ the context for $2.8$ points. At that volume the
feature stops being a retrieval setting and becomes a different product, one
whose per-query cost most callers will not accept. The capability is built and
it functions; what is missing is a policy for using it sparingly.

More context is also not monotonically better. Three question types score
\emph{worse} under C than under B: \texttt{temporal\_reasoning}
($56.1 \rightarrow 47.3$, comparing the single B run that C was measured against
rather than the five-run mean of 54.8 in Table~\ref{tab:bytype}),
\texttt{knowledge\_update}, and
\texttt{contradiction\_resolution}. Those are the three types where the answer
depends on distinguishing between similar records, and expansion supplies more
similar records. This points at the open question directly: not how much context
to deliver, but to which questions.

The run is incomplete as well: $70.3\%$ is over the 681 questions that
completed, against 700 for B.

We therefore report C as a preliminary figure and not as our result. It says
headroom exists above B and that reaching it needs work we have not done: a
policy for spending expansion on the questions that repay it, and a complete
run. Both are open.

\subsection{Where we score badly, and why}

Figure~\ref{fig:bytype} draws the by-type scores.
\texttt{event\_ordering} at \pct{23.6} is our worst type by a wide margin. We think it measures something other than ordering ability. The rubric expects ten
\emph{topics} (``translation API integration and error handling''); our answers
give ten \emph{individual events} (``March 12, asked how to cut franc from
100\,ms to 50\,ms''). Both are in chronological order and both have ten items,
but they sit at different levels of granularity, so the equivalence matcher
fails to pair them. Unpaired items are pushed to the bottom of the ranking and
the correlation goes negative, below the level of a random ordering.

We could raise this score by reshaping our answers to match the rubric's
granularity. We did not, because that is answering to the answer key rather than
to the question; the benchmark's instruction says only to mention exactly ten
items and does not specify the unit.

\paragraph{Corpus defects.} 211 turns (0.3\%) longer than 8{,}000 characters were
truncated by our ingest harness, and one turn of 74{,}630 failed to store. Both
are our harness's fault, not the benchmark's, and both remain in the measured
corpus.

\paragraph{Judge cost.} Nugget scoring calls the judge once per rubric item, an
average of 3.45 times per question, and \texttt{event\_ordering} pairs every
answer line against every rubric item. Five runs required 38{,}662 judge calls,
25.3M tokens, roughly \$12, more than the question count would suggest.

\FloatBarrier
\section{Benchmark 3: multilingual retrieval of photographs}
\label{sec:multimodal}

\subsection{Why a lexical control is the whole point}

``Language-agnostic retrieval'' is cheap to assert. A system can report high
accuracy in ten languages and the reader still cannot tell whether the system is
good or the task is easy. What settles it is a baseline that is known to be
language-\emph{dependent}, run on the same corpus with the same queries.

BM25 \citep{robertson2009bm25} is that baseline. It scores a query against a
document by term overlap, so it works when the query and the document are
written in the same language and fails when they are not. The interesting part
is not that it fails. What matters is where and how much, because that
gives the shape of the dependence rather than a slogan.

There is a condition where the comparison stops being quantitative. A photograph
stored with no caption has no document text. BM25 does not score it badly; there
is nothing to score. Any retrieval of that photograph from a text query is
outside what lexical matching can do at all. We use that condition as the
central measurement, and report the quantitative cells around it.

\paragraph{We configured BM25 to be as strong as we could make it.}
This determines whether the section is worth anything, so we state it in full.
Okapi BM25 with standard $k_1=1.5$, $b=0.75$. Two tokenisers were run over every
cell: whitespace with case folding, and character bigrams, the standard fallback
for scripts that do not delimit words, and \emph{the better of the two is
reported per cell}. Choosing one tokeniser globally would have handicapped
either the space-delimited or the non-space-delimited languages, and beating a
badly tokenised baseline would prove nothing about BM25.

The evidence that this worked: every diagonal cell, where query language matches
caption language, comes out at exactly \pct{100}. A weakened baseline fails there
first. Our implementation is dependency-free and included in the release, so the
BM25 half of this section can be reproduced with no access to our engine.

\subsection{Controlled corpus}
\label{sec:controlled}

Thirty images, five colours by six shapes, generated so that the correct answer
for a query like ``red circle'' is unique and a human can verify the ranking by
eye. Seven store conditions crossed with ten query languages gives 70 cells,
each of 30 queries.

The store conditions matter more than they look. In a store where
\emph{some} photographs have captions, the captioned ones crowd the five
available result slots and push caption-less answers out. Our first version of this experiment mixed them and produced the confident
and wrong conclusion that text queries cannot find caption-less photographs at
all (\pct{0}). Separating the stores reversed it. We report the mixed condition
as a result rather than hiding it, because real customer stores are mixed.

\begin{table}[t]
\centering
\small
\caption{Controlled corpus, recall@5 (\%). 70 cells $\times$ 30 queries.
Query languages: Korean, English, Japanese, Chinese, Arabic, Russian, Hindi,
Spanish, German, Thai. Rows are what the store holds.}
\label{tab:matrix}
\begin{tabular}{lrrrrrrrrrr}
\toprule
Store holds & ko & en & ja & zh & ar & ru & hi & es & de & th \\
\midrule
\multicolumn{11}{l}{\emph{BM25}}\\
Korean captions   & \hi{100} & \zero{0} & \zero{0} & \zero{0} & \zero{0} & \zero{0} & \zero{0} & \zero{0} & \zero{0} & \zero{0} \\
English captions  & \zero{0} & \hi{100} & \zero{0} & \zero{0} & \zero{0} & \zero{0} & \zero{0} & 53 & 67 & \zero{0} \\
Japanese captions & \zero{0} & \zero{0} & \hi{100} & 43 & \zero{0} & \zero{0} & \zero{0} & \zero{0} & \zero{0} & \zero{0} \\
Chinese captions  & \zero{0} & \zero{0} & 43 & \hi{100} & \zero{0} & \zero{0} & \zero{0} & \zero{0} & \zero{0} & \zero{0} \\
Arabic captions   & \zero{0} & \zero{0} & \zero{0} & \zero{0} & \hi{100} & \zero{0} & \zero{0} & \zero{0} & \zero{0} & \zero{0} \\
All five languages& \hi{100} & \hi{100} & \hi{100} & \hi{100} & \hi{100} & \zero{0} & \zero{0} & 53 & 70 & \zero{0} \\
\textbf{No caption} & \zero{0} & \zero{0} & \zero{0} & \zero{0} & \zero{0} & \zero{0} & \zero{0} & \zero{0} & \zero{0} & \zero{0} \\
\midrule
\multicolumn{11}{l}{\emph{\sys{}}}\\
Korean captions   & 100 & 93 & 97 & 93 & 90 & 97 & 83 & 87 & 90 & 83 \\
English captions  & 100 & 100 & 100 & 100 & 100 & 100 & 70 & 100 & 100 & 93 \\
Japanese captions & 90 & 100 & 100 & 100 & 90 & 100 & 83 & 100 & 100 & 87 \\
Chinese captions  & 100 & 100 & 100 & 100 & 100 & 100 & 97 & 100 & 100 & 97 \\
Arabic captions   & 97 & 87 & 93 & 90 & 100 & 93 & 80 & 90 & 97 & 83 \\
All five languages& 100 & 100 & 100 & 100 & 100 & 100 & 93 & 100 & 97 & 93 \\
\textbf{No caption} & 100 & 100 & 100 & 100 & 97 & 100 & 70 & 100 & 100 & 87 \\
\bottomrule
\end{tabular}
\end{table}

\begin{figure}[p]
\centering
\includegraphics[width=\textwidth]{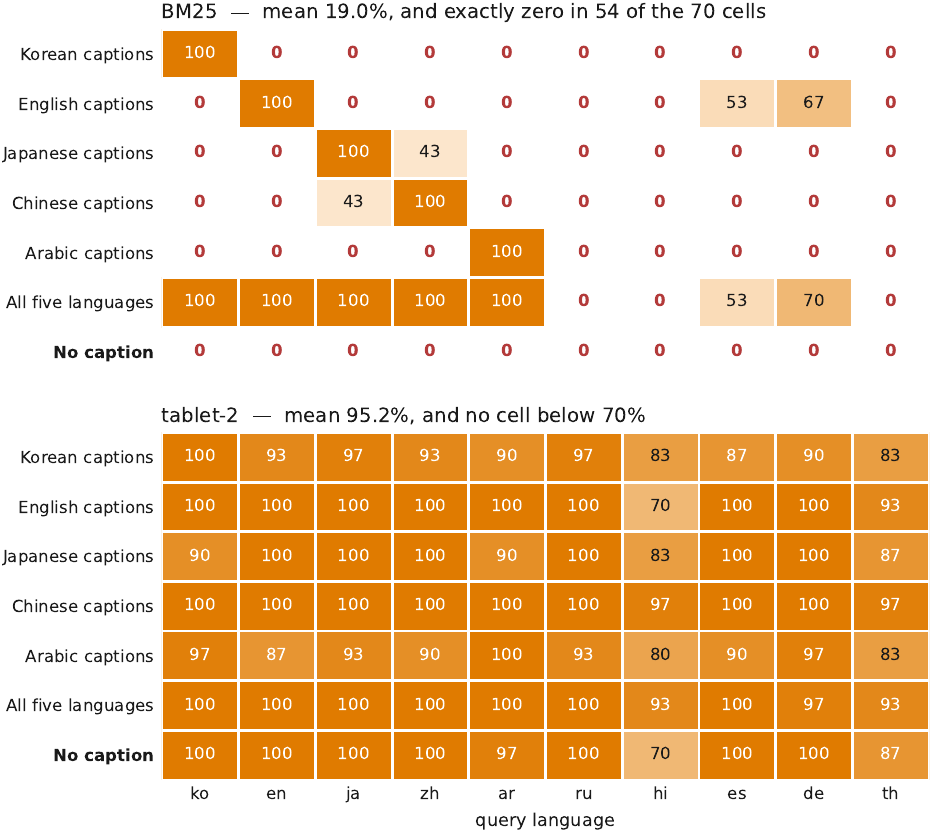}
\caption{Table~\ref{tab:matrix} shaded. Every value is the same; the shading is
there so that the shape of the lexical dependence is visible without reading
seventy numbers. BM25's non-zero cells form a diagonal plus two off-diagonal
pairs, and those pairs are exactly the language relationships a tokeniser can
see: Japanese and Chinese share Han characters, and English shares cognates
with German and Spanish. The bottom row of each panel is the condition that
carries the argument.}
\label{fig:matrix}
\end{figure}

\begin{table}[t]
\centering
\caption{Aggregate over the 70 cells of Table~\ref{tab:matrix}.}
\label{tab:matrixagg}
\begin{tabular}{lrr}
\toprule
recall@5 (\%) & BM25 & \sys{} \\
\midrule
Mean over 70 cells      & 19.0 & 95.2 \\
Median                  & 0.0  & 100.0 \\
Cells scoring exactly 0 & 54 of 70 & none \\
Mean where the store held no caption in the query's language (60 cells) & 5.5 & 94.2 \\
Mean excluding only the matching-language diagonal (65 cells) & 12.8 & 94.7 \\
\bottomrule
\end{tabular}
\end{table}

Table~\ref{tab:matrixagg} aggregates the 70 cells and
Figure~\ref{fig:matrix} shades them. Three things are visible in
Table~\ref{tab:matrix}.

\paragraph{Lexical transfer happens only between relatives.}
The only non-zero off-diagonal BM25 cells are Japanese$\leftrightarrow$Chinese
(43\%), where the two scripts share Han characters outright, and
English$\rightarrow$German (67\%) and English$\rightarrow$Spanish (53\%), where
cognates give the bigram tokeniser partial overlap. This is a sharper statement
than ``BM25 is not multilingual'': it transfers exactly as far as orthographic
and etymological relatedness reaches, and no further.

\paragraph{Lexical retrieval can be made flat, by paying at write time.}
The ``all five languages'' row is a store where the operator anticipated every
language and wrote every caption five times. BM25 reaches \pct{100} on all five.
It also reaches \pct{0} on Russian, Hindi and Thai, which were not bought. This
is the honest form of the argument: multilingual lexical search is achievable,
at the cost of enumerating the languages in advance and re-indexing when a
sixth arrives. Dense retrieval does not have a language list.

\paragraph{The bottom row is the one that matters.}
Ten languages, all exactly zero, because there is no document. \sys{} ranges
from \pct{70} to \pct{100} in the same condition.

\subsection{Real photographs: Crossmodal-3600}

Synthetic shapes invite the objection that the task was easy and that we wrote
the queries. Crossmodal-3600 \citep{thapliyal2022xm3600} answers both. It is
3{,}600 geographically diverse photographs with captions in 36 languages, and
crucially the captions are \emph{not translations}: a native speaker of each
language wrote them while looking at the photograph. That distinction is the
reason we use this corpus and not a larger one. The usual way to obtain a
multilingual evaluation set is to translate an existing English one, as mMARCO
does for passage ranking \citep{bonifacio2021mmarco}, and a translated query
inherits the English original's phrasing, so retrieving against it measures
translation quality alongside retrieval. Here each language describes the
photograph in its own terms, which is both harder and closer to what a caller
actually sends.

We sampled 300 photographs having captions in all fourteen languages we chose,
using a fixed seed. The languages span eight scripts and a wide range of
resource levels, deliberately including Swahili, Telugu and Hindi. Each
photograph is stored with no caption; the query is that photograph's
human-written caption in the target language; the task is to find it among 300.
Random guessing gives \pct{0.33} at rank 1 and \pct{1.7} at rank 5.

\begin{table}[t]
\centering
\small
\caption{Crossmodal-3600, 300 photographs stored with no caption. \sys{} figures
are measured on the current build; brackets are Wilson 95\%
intervals. BM25 has no document to score in this condition, the stored records
containing no text.}
\label{tab:xm}
\begin{tabular}{llrrr}
\toprule
Language & Script & recall@1 (\%) & recall@5 (\%) & BM25 recall@5 \\
\midrule
Russian    & Cyrillic   & 90.7 & 100.0 \,[98.7--100.0] & \zero{0.0} \\
Japanese   & Kana/Han   & 92.0 & 99.7 \,[98.1--99.9]  & \zero{0.0} \\
German     & Latin      & 93.7 & 99.3 \,[97.6--99.8]  & \zero{0.0} \\
Chinese    & Han        & 87.0 & 99.3 \,[97.6--99.8]  & \zero{0.0} \\
Spanish    & Latin      & 86.7 & 99.0 \,[97.1--99.7]  & \zero{0.0} \\
Vietnamese & Latin      & 87.7 & 97.7 \,[95.3--98.9]  & \zero{0.0} \\
English    & Latin      & 82.0 & 97.0 \,[94.4--98.4]  & \zero{0.0} \\
Korean     & Hangul     & 87.3 & 97.0 \,[94.4--98.4]  & \zero{0.0} \\
Arabic     & Arabic     & 81.3 & 96.3 \,[93.6--97.9]  & \zero{0.0} \\
Hebrew     & Hebrew     & 86.7 & 95.7 \,[92.7--97.5]  & \zero{0.0} \\
Thai       & Thai       & 86.3 & 95.7 \,[92.7--97.5]  & \zero{0.0} \\
Hindi      & Devanagari & 58.3 & 86.0 \,[81.6--89.5]  & \zero{0.0} \\
Telugu     & Telugu     & 40.3 & 64.0 \,[58.4--69.2]  & \zero{0.0} \\
Swahili    & Latin      & 26.7 & 53.0 \,[47.3--58.6]  & \zero{0.0} \\
\midrule
\multicolumn{2}{l}{Mean} & 77.6 & 91.4 & 0.0 \\
\bottomrule
\end{tabular}
\end{table}

\begin{figure}[t]
\centering
\includegraphics[width=\textwidth]{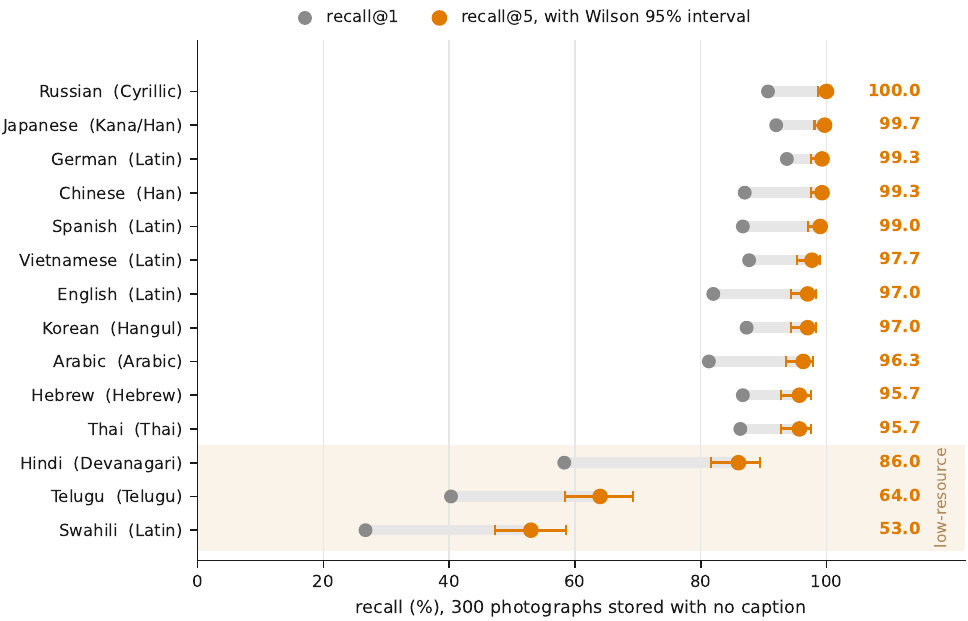}
\caption{Table~\ref{tab:xm} plotted, with the recall@1 and recall@5 values for
each language joined so that the distance between them is visible. Eight
scripts sit above \pct{95}, which is the case against reading the spread as a
script effect. The three shaded rows are the low-resource languages, and they
fall away in the same order regardless of what they are written in: Swahili is
Latin script and scores worst.}
\label{fig:xm}
\end{figure}

Figure~\ref{fig:xm} draws the same table.
For the quantitative comparison we repeat the measurement with an English
caption attached to every photograph, which is the most favourable realistic
setting for a lexical system. BM25 then scores \pct{100} on English, where the query
and the document are the same language, written by the same annotator, and at
most \pct{9.7} on any of the other thirteen: German 9.7, Spanish 8.0, Swahili
5.0, Vietnamese 3.0, Russian 0.7, Thai 0.3, Korean 0.3, and exactly zero for
Arabic, Hebrew, Hindi, Telugu, Japanese and Chinese.

\subsection{Open baselines: would anything that avoids term matching do?}
\label{sec:densebaseline}

The lexical control establishes that term matching cannot do this task. It does not establish that \emph{we} can. A reader may reasonably suspect that any system built for image retrieval would clear the same bar. That objection deserves a
measurement rather than an argument, so we ran two open models over the same 300
photographs and the same human-written captions.

The baseline had to be assembled rather than cited. Multilingual retrieval is
well served by existing benchmarks, but they are text against text: Mr.\ TyDi
\citep{zhang2021mrtydi} and MIRACL \citep{zhang2023miracl} score passage
retrieval across eighteen languages, and BEIR \citep{thakur2021beir} and MTEB
\citep{muennighoff2023mteb} aggregate suites for the encoders that do it. None
of them contains a document with no text in it, which is the condition this
section exists to test.

The two models we ran are chosen to separate a capability from its training
data. Both use the same image encoder \citep{radford2021clip}; only the text side
differs. One pairs it with that model's original English-trained text encoder,
the other with a text encoder distilled to cover more than fifty languages
\citep{reimers2020multilingual}. Whatever changes between the two rows is therefore
the text side alone. Newer open image-text models would place the level
comparison better \citep{zhai2023siglip}, and we did not run one, because a
different model brings a different image encoder with it: the difference between
rows would then mix the vision side back in, which is the one thing this
measurement is built to hold still.

\begin{table}[t]
\centering
\small
\caption{Recall@5 (\%) on the same 300 caption-less photographs and the same
queries. The two open baselines share an image encoder and differ only in their
text encoder. BM25 is included as a reminder that in this condition it is not
low but undefined.}
\label{tab:dense}
\begin{tabular}{lrrrr}
\toprule
Language & CLIP, English text tower & CLIP, multilingual text tower & \sys{} & BM25 \\
\midrule
English    & 91.0 & 85.7 & 97.0 & \zero{0.0} \\
Spanish    & 66.0 & 87.0 & 99.0 & \zero{0.0} \\
German     & 49.7 & 83.0 & 99.3 & \zero{0.0} \\
Japanese   & 22.0 & 86.0 & 99.7 & \zero{0.0} \\
Chinese    &  8.7 & 87.3 & 99.3 & \zero{0.0} \\
Vietnamese &  8.3 & 88.3 & 97.7 & \zero{0.0} \\
Thai       &  5.3 & 71.0 & 95.7 & \zero{0.0} \\
Swahili    &  6.0 &  6.0 & 53.0 & \zero{0.0} \\
Russian    &  4.7 & 88.3 & 100.0 & \zero{0.0} \\
Korean     &  3.0 & 71.7 & 97.0 & \zero{0.0} \\
Hebrew     &  2.3 & 75.7 & 95.7 & \zero{0.0} \\
Arabic     &  2.0 & 74.7 & 96.3 & \zero{0.0} \\
Hindi      &  2.0 & 48.7 & 86.0 & \zero{0.0} \\
Telugu     &  1.0 &  5.0 & 64.0 & \zero{0.0} \\
\midrule
Mean       & 19.4 & 68.5 & 91.4 & 0.0 \\
Median     &  5.7 & 79.3 & 97.0 & 0.0 \\
Std.\ dev. & 27.5 & 27.7 & 14.0 & --- \\
\bottomrule
\end{tabular}
\end{table}

\begin{figure}[t]
\centering
\includegraphics[width=\textwidth]{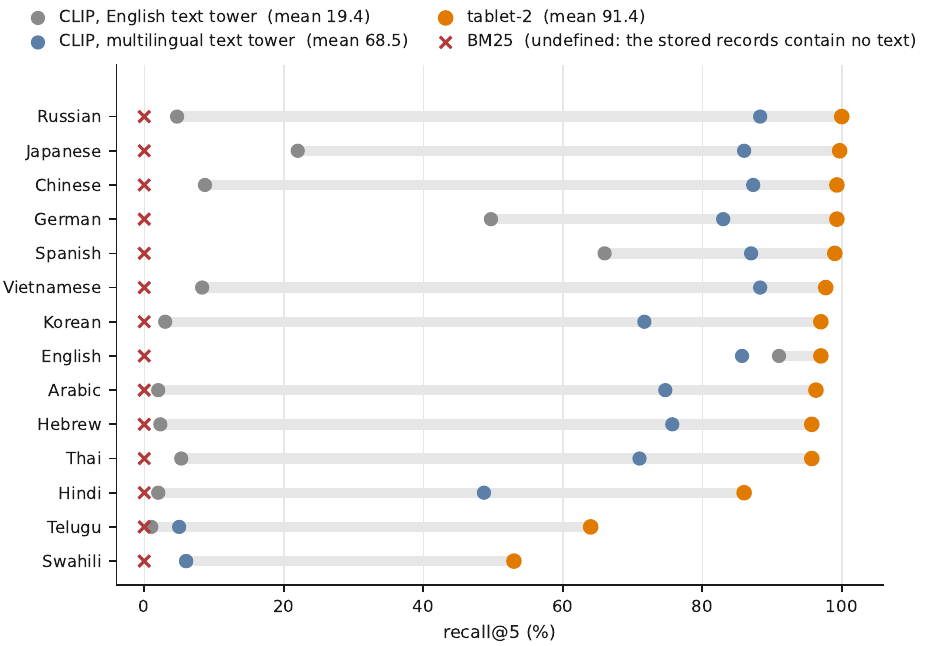}
\caption{Table~\ref{tab:dense} ordered by our own score. The connecting bar runs
from the English-only text tower to us, so its length is the range the same
image vectors can be made to produce by changing nothing but the text side.
The claim this figure supports is about shape rather than level: the baselines
are a 2021 open architecture at base scale and \sys{} is a current production
system (\S\ref{sec:densebaseline}). What is not a level effect is that both
baselines have a language they collapse on and we do not.}
\label{fig:dense}
\end{figure}

Figure~\ref{fig:dense} orders the same data by our own score.

\paragraph{Being dense does not make a system language-agnostic.}
The English text tower reaches \pct{91.0} on English and \pct{4.7} on Russian.
Both rows retrieve from identical image vectors, so this is not a vision
failure; it is a text encoder that was never asked to cover those languages.
Dense retrieval is not automatically flat across languages, and a paper claiming
flatness has to show it rather than assume it follows from not using BM25.

\paragraph{A multilingual text tower is flat only over the languages it was
trained on.}
Swapping in the multilingual encoder lifts the mean from \pct{19.4} to
\pct{68.5}, and across nine of the fourteen languages it is genuinely level, in
the low to high eighties. Then it falls off a cliff: Telugu \pct{5.0} and
Swahili \pct{6.0}, no better than the English-only tower. This is the same
shape as the lexical result in \S\ref{sec:controlled}, where writing captions in
five languages made BM25 flat over exactly those five and left the rest at zero.
The coverage is bought in advance in both cases; only the currency differs.
This is how coverage works throughout the family of multilingual text encoders
this baseline belongs to. It is decided at training time and published as a
count: more than 100 working languages for one \citep{chen2024m3}, 100 for
another, inherited from the model it was initialised from \citep{wang2024me5},
and bitext retrieval reported over 112 for a third \citep{feng2022labse}. The
authors of the second add that low-resource languages among their hundred may
still degrade, which is the distinction that matters here. Being on the list is
not the same as being served by it, and a measurement in fourteen languages is
partly a measurement of whose list they are on and where on it they sit.

\paragraph{Where we are weakest is where the field is weakest, by an order of
magnitude.}
Section~\ref{sec:limits} reports Swahili at \pct{53.0} and Telugu at \pct{64.0}
as our worst results, and they remain our worst results. Against the strongest
open baseline available to us they are \pct{6.0} and \pct{5.0}. Averaged over
the three low-resource languages, the multilingual baseline reaches \pct{19.9}
and we reach \pct{67.7}. We report our weakness as a weakness, and note that on
this corpus no open system we could run does better. Swahili and Telugu are both
among MIRACL's eighteen languages \citep{zhang2023miracl}, so the field does
track them for text; what it does not track is finding a photograph in them.

\paragraph{What this comparison is not.}
The open models here are a widely used 2021 architecture at base scale, and a
large part of the absolute gap is that. We are not claiming to have out-built
them, and this measurement could not establish it either
way. The comparison is informative about \emph{shape} rather than level: the
standard deviation across languages is 27.5 and 27.7 for the two baselines
against 14.0 for us, and the two baselines lose 90.0 and 83.3 points
respectively between their best and worst language, against 47.0 for us.
A reader who wants a level-matched comparison would need a current commercial
system on the baseline side, which we could not obtain.

Note also that the multilingual tower scores \emph{lower} on English than the
English-only tower it replaced, \pct{85.7} against \pct{91.0}. Coverage was
bought with a little of the language it started from.

\subsection{Negative result: captions make cross-lingual retrieval worse}
\label{sec:capharm}

Attaching those English captions did not only help English.

\begin{table}[t]
\centering
\caption{Effect of attaching an English caption to every photograph.
\sys{} recall@5 (\%), same 300 photographs.}
\label{tab:capharm}
\begin{tabular}{lrrr}
\toprule
& No caption & English caption & $\Delta$ \\
\midrule
English            & 97.0 & 100.0 & $+3.0$ \\
Japanese           & 99.7 & 92.3 & $-7.4$ \\
Chinese            & 99.3 & 86.7 & $-12.6$ \\
Korean             & 97.0 & 85.7 & $-11.3$ \\
Hindi              & 86.0 & 69.7 & $-16.3$ \\
Swahili            & 53.0 & 33.0 & $-20.0$ \\
\midrule
Mean over 14 languages & 91.4 & 80.0 & $-11.4$ \\
\bottomrule
\end{tabular}
\end{table}

\begin{figure}[t]
\centering
\includegraphics[width=\textwidth]{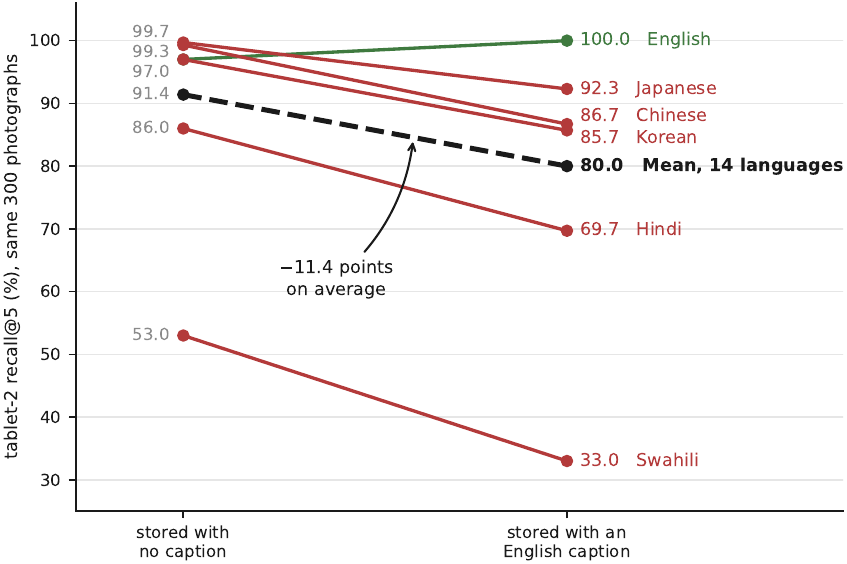}
\caption{Attaching an English caption to every photograph, drawn as a slope so
that the single line going the other way is easy to find. English gains three
points; the other thirteen lose eleven on average, and the loss is largest
where our score was already lowest. This is a defect in a shipping product
rather than a property of the benchmark: customer stores are mixed in exactly
this way.}
\label{fig:capharm}
\end{figure}

Table~\ref{tab:capharm} and Figure~\ref{fig:capharm} give the effect per
language. One language gains three points and thirteen lose an average of
eleven. The
caption helps a query written in its own language and gets in the way of every
other, and on this corpus the second effect is the larger one. We report the
measurement; the reason it happens is a property of our retrieval that we are
not describing here.

This reproduces, at scale and on real photographs, the ``mixed store'' effect we
saw on the synthetic corpus. It is a defect in our product, not a property of
the benchmark: real customer stores contain a mixture of captioned and
caption-less images, and a photograph that a user could have found is being
pushed out of the top five by its neighbours' captions. We have not fixed it.
Making a stored caption help the queries it can help without penalising the
rest is open work.

\subsection{Uncertainty}

Three independent re-ingests of the same corpus, queried identically, gave
recall@5 of 86.7, 83.3 and 90.0 ($\sigma=2.7$), and recall@1 of 36.7, 33.3, 33.3
($\sigma=1.6$). The Wilson 95\% interval for a single one of those measurements
at $n=30$ is $[70.3, 94.7]$, about $\pm12$ points.

We report that comparison rather than a run count because it is the honest
summary: at $n=30$ the uncertainty is dominated by corpus size, not by
run-to-run variation, and reporting ``averaged over five runs'' would suggest a
precision the corpus cannot support. The 300-photograph measurements have
intervals of roughly $\pm4$ points and we give them per cell in
Table~\ref{tab:xm}.

\FloatBarrier
\section{Attribution: whose limit is a weak language?}
\label{sec:attrib}

The tables above span a wide range across languages, from \pct{100} in Russian
to \pct{53} in Swahili. A system paper is entitled to say ``that is the
component, not us,'' and we said exactly that once and were wrong, and said it a
second time and were right. The procedure that separated the two cases is the
part of this section worth carrying away.

\paragraph{Procedure.}
Take the rest of the system out of the measurement. Put the same photographs and
the same queries through that one stage, score every photograph directly, order
by the result, and compare against what the full service returned. If the two
agree, the handling around the stage is adding no loss and the language profile
belongs to the stage. If the full service is worse, the loss is in the handling,
and it is findable.

\subsection{Case 1: a limit of ours that we had called the component's}

On the synthetic corpus, Korean recall@1 was \pct{33.3} while nine other
languages tracked the component closely. We had recorded the conclusion that
Korean colour vocabulary was simply handled poorly and that nothing could be
done about it, the component being the current version already.

Table~\ref{tab:inputtype} sets the configurations side by side. Measured on its
own, that stage reached \pct{73.3} on the same task. The gap was
in our handling. The cause was one setting on the way in, which tells the stage
whether an item is being stored or looked for. Our text path had set it
correctly since the beginning. Our photograph path did not set it at all and
could not: the field was absent from the request type, so no code could have
supplied it.

\begin{table}[t]
\centering
\caption{Synthetic corpus, Korean, 30 photographs stored with no caption. The setting omitted reproduces the engine's measured value exactly, and that
match is what identifies the cause. Nine other languages moved by less
than 5 points under every configuration.}
\label{tab:inputtype}
\begin{tabular}{lrr}
\toprule
Configuration & recall@1 (\%) & recall@5 (\%) \\
\midrule
Setting omitted \emph{(what the engine did)} & 33.3 & 90.0 \\
Set on the incoming query only & 70.0 & 100.0 \\
Set on both sides & 73.3 & 100.0 \\
\midrule
Engine, measured, before the change & 33.3 & 90.0 \\
Engine, measured, after the change  & 70.0 & 100.0 \\
\bottomrule
\end{tabular}
\end{table}

We changed it on the incoming-query side only. Changing it on both is marginally
better in isolation, but every photograph already held was written under the old
setting, and mixing two conventions in one store would misrank them until all of
them had been written again. The measured difference between the two options is
one photograph in thirty.

Two things about this are worth recording. The effect is confined to short
queries: on the real-photograph corpus, where a query is a full human-written
caption, the same change is worth $+1.3$ points at rank 1 on average and not
$+37$. A long query carries enough signal that one request field cannot reorder
the top of the list. And we spent time on a wrong hypothesis first. Our first
candidate was a configuration default that would have cost Korean, whose score
margins here are compressed, more than English, whose are wider, and it fitted
the symptom exactly. One run ruled it out. A hypothesis that explains the data
is not the cause until it has been tested, and one that fits this well is the
kind that gets believed instead.

\subsection{Case 2: low-resource languages, which really are the component's}

The same procedure applied to all fourteen languages on the real-photograph
corpus returns the opposite verdict.

\begin{table}[t]
\centering
\caption{The full service versus that one stage measured on its own, same 300
caption-less photographs, same queries, recall@5 (\%). \textbf{Both columns
predate the change in \S\ref{sec:attrib}}, which is what makes them comparable:
at that point both used the same request convention, so the only difference left
between them is our own retrieval path. Table~\ref{tab:xm} reports the engine
after the change and is 0 to 6 points higher.}
\label{tab:attrib}
\begin{tabular}{lrrr}
\toprule
Language & Full service & Stage alone & Difference \\
\midrule
Russian    & 100.0 & 100.0 & $0.0$ \\
Japanese   & 99.7 & 99.7 & $0.0$ \\
German     & 99.3 & 99.3 & $0.0$ \\
Spanish    & 99.0 & 99.0 & $0.0$ \\
Vietnamese & 98.0 & 98.0 & $0.0$ \\
Chinese    & 98.0 & 98.0 & $0.0$ \\
Korean     & 97.7 & 97.0 & $+0.7$ \\
English    & 97.0 & 97.0 & $0.0$ \\
Arabic     & 95.3 & 95.7 & $-0.4$ \\
Hebrew     & 95.3 & 95.3 & $0.0$ \\
Thai       & 94.7 & 95.3 & $-0.6$ \\
Hindi      & 83.3 & 83.3 & $0.0$ \\
Telugu     & 64.3 & 64.3 & $0.0$ \\
Swahili    & 47.3 & 47.0 & $+0.3$ \\
\bottomrule
\end{tabular}
\end{table}

\begin{figure}[t]
\centering
\includegraphics[width=\textwidth]{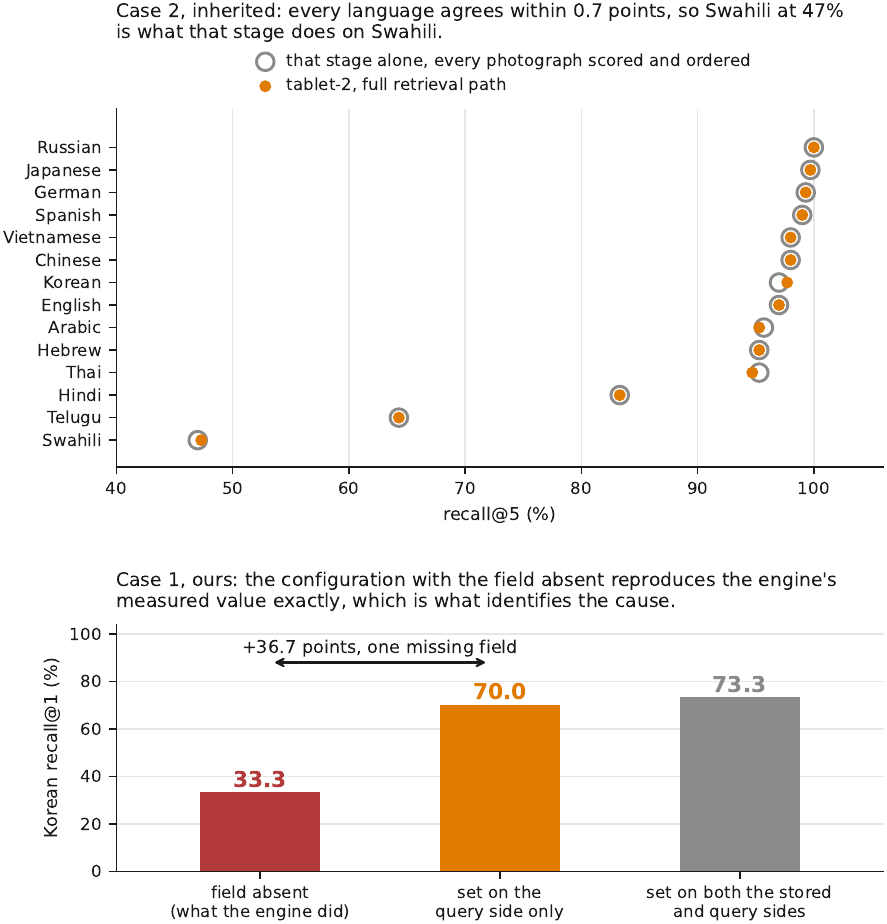}
\caption{The same procedure, run twice, returning a different verdict each time.
Where the two bars meet, the full service is at the ceiling that stage
supports; where they part, the gap was in the handling and ours to fix.}
\label{fig:attrib}
\end{figure}

Table~\ref{tab:attrib} gives all fourteen and Figure~\ref{fig:attrib} draws the
two cases together. Every language agrees within 0.7 points. Swahili at \pct{47} is what that stage
can do on Swahili, and the same holds for Telugu and Hindi.

The reading that matters here is not that our retrieval adds nothing. The
stage-alone column is produced by scoring every photograph in the corpus and
ordering by that score, which is the best any system built on that stage could
do. It is the ceiling and not a baseline. \sys{} reaches it without examining
every record, and while
doing the rest of what a memory service does around it. Matching exhaustive
scoring under those conditions is the result. It also bounds what we can promise: cross-lingual coverage moves when that
stage moves and not otherwise.

\begin{table}[t]
\centering
\caption{Two versions of that stage, each measured on its own on the synthetic
corpus, reported as earlier and current.}
\label{tab:stagever}
\begin{tabular}{lrrrr}
\toprule
& \multicolumn{2}{c}{recall@1 (\%)} & \multicolumn{2}{c}{recall@5 (\%)} \\
\cmidrule(lr){2-3}\cmidrule(lr){4-5}
Language & earlier & current & earlier & current \\
\midrule
Korean & 40.0 & 73.3 & 86.7 & 100.0 \\
Hindi  & 0.0  & 20.0 & 36.7 & 73.3 \\
Arabic & 36.7 & 56.7 & 90.0 & 96.7 \\
Thai   & 60.0 & 60.0 & 83.3 & 90.0 \\
\bottomrule
\end{tabular}
\end{table}

Table~\ref{tab:stagever} supports that last point directly: a version change in
that stage moves the whole language profile, with the largest gains where the
profile was weakest, Hindi recall@5 going from \pct{36.7} to \pct{73.3}.

\FloatBarrier
\section{Limitations}
\label{sec:limits}

\paragraph{Every number here is self-reported, including the comparisons.}
We ran our own benchmarks on our own engine. So did every system in
Table~\ref{tab:compare}. We used each benchmark's default judge without changing
it, which is the most defensible choice available to us, but it is not
equivalent to knowing that other reports did the same.

\paragraph{The judge is a model.}
Both benchmarks decide correctness with a language model, which is the field's
standard \citep{zheng2023judge} and not a neutral instrument: judges are
sensitive to how candidates are presented to them, and in particular to the
order in which two are shown \citep{wang2024unfair}. Our exposure to that
specific failure is limited, since neither protocol here places two answers side
by side; each verdict is one answer against a reference. The ordinary exposure
remains. A judge wrong in a consistent direction moves every configuration in
this paper by the same amount, which leaves the comparisons intact and the
absolute levels not. The released grade files carry each question's score with
the number of rubric items behind it, and the harness re-grades from a recorded
run without repeating retrieval, so a reader who doubts a verdict can obtain
their own rather than argue with ours.

\paragraph{The reader is not controlled across systems.}
This is the largest single caveat on the text results and we have quantified
part of it: 2.0 points from the reader model on LongMemEval-S, 8.9 from the
re-ask budget on BEAM-1M. Those are lower bounds on how much a comparison across
differently-configured systems can move.

\paragraph{Recall@5 is a ceiling, not a choice.}
A single response carries at most five photographs. Every multimodal figure in
this paper is therefore measured against that limit, and ``not in the top five''
is the floor of what these experiments can observe. A system that places the
right photograph at rank six is indistinguishable here from one that never
finds it.

\paragraph{Corpus size dominates the multimodal uncertainty.}
The synthetic corpus has 30 images. Three independent re-ingests varied by
$\sigma=2.7$ points, while the binomial interval at $n=30$ is about $\pm12$. The
300-photograph measurements are tighter, at roughly $\pm4$, and we give per-cell
intervals for them.

\paragraph{Cross-language comparison on XM3600 is not like-for-like.}
Its captions are written independently by native speakers rather than
translated, which is what makes the benchmark honest about language, and also
means a Korean caption and an English caption of the same photograph describe
different things. Comparing our Korean number against our English number
therefore mixes retrieval quality with what each group of annotators chose to
mention. The comparison this paper relies on is within a language, between
lexical retrieval and ours.

\paragraph{The low-resource weakness is real, and is shared.}
Swahili at \pct{53.0} and Telugu at \pct{64.0} recall@5 are our worst results
and we do not present them otherwise. Section~\ref{sec:densebaseline} adds the
context that the strongest open baseline we could run scores \pct{6.0} and
\pct{5.0} on the same two languages, so the deficiency is not peculiar to us. It
is still a deficiency, and a caller serving those languages should size their
expectations by these numbers rather than by our mean.

\paragraph{The dense baselines are not level-matched.}
They are a 2021 open architecture at base scale, run against a system in current
production use. The absolute gap is therefore not a claim about our engineering and we make none. What the comparison supports is the narrower claim about shape: that
retrieving densely does not by itself produce language independence, and that a
system which is level across languages is level because someone made it so
and not as a consequence of not using term matching. A level-matched
comparison would need a current commercial system on the baseline side, which we
could not obtain.

\paragraph{Cross-lingual coverage is inherited.}
Section~\ref{sec:attrib} shows our retrieval matching, to within 0.7 points on
all fourteen languages, what exhaustive scoring of the same corpus achieves.
Read one way that is the ceiling and we are at it. Read another it means the
multilingual result is largely a property of that one stage and moves when the
stage does. Both readings are correct, and the second is why these numbers
describe \sys{} as measured in August 2026.

\paragraph{Known defects we have not fixed.}
Captions attached to photographs degrade retrieval for queries in other
languages by 11.4 points on average (\S\ref{sec:capharm}); real customer stores
are mixed in exactly this way. Nineteen BEAM questions errored during
configuration C and have not been re-run. Of the 74{,}630 ingested turns, 211
longer than 8{,}000 characters were truncated by our harness and one failed to
store.

\FloatBarrier
\section{Reproducibility}
\label{sec:repro}

The text benchmarks were run with a harness we release, which takes a task
file, a retrieval endpoint, a reader command and a judge command, and records
every question with the context it received. Two of its properties bear on the numbers here. It counts tokens only when given a tokeniser command and
never estimates them, which is the check that eventually caught our own error in
\S\ref{sec:tokens}. And it refuses to grade a run whose answers look like
provider errors, because a run whose reader was failing produces a plausible
number describing nothing.

It grades two protocols, and choosing between them is not a
detail. The binary one gives a single yes-or-no verdict per question, which is
how LongMemEval scores. BEAM-1M does not: its questions carry rubric items,
each judged on a three-point scale, with the question's score their mean, and
its ordering questions scored by rank correlation rather than by that mean. Both
protocols are in the released harness and either will print a number for either
benchmark. Grading BEAM the wrong way moves the total by roughly four points,
which is larger than the spread between the systems in
Table~\ref{tab:compare}, so the harness requires the protocol to be named
and does not guess it from the task file.

The rank correlation is implemented in plain Python rather than imported, so the
harness runs with no third-party packages. We checked it against the standard
library implementation on 400 random inputs with ties and found no
disagreement, and the released aggregation reproduces all five of our runs to
within 0.05 points.

\paragraph{Access.}
\sys{} was in preview while this paper was written and is on general release
from 25 August 2026. Everything in this section runs against our earlier
generations at any time, and against \sys{} itself from that date. We state it
because a reader who reaches for the model named in the title should learn where
it stands from the paper and not from an error.

The harness, the scripts that build the task file and load and verify the
corpus, and the per-question record of every configuration reported here are
published at \url{https://github.com/wontopos/beam1m-tablet-2} on 29 August 2026,
including the two we do not present as our result: the re-ask-off run and the
expansion run that lost nineteen questions to errors. We give the date rather
than a bare link because a reader who follows it before then should know they
are early, not that we misplaced it. Records are keyed by each benchmark's own question
identifier, and the questions, ground truths and rubrics are not reproduced
there, because they belong to the benchmark authors and are obtained from them.
A reader who has the benchmark can therefore recompute any figure we quote from
these runs rather than take it.

The BM25 control has no dependency on our engine at all. It is Okapi BM25 with
$k_1=1.5$, $b=0.75$ over the caption strings, implemented in plain Python with no
third-party imports.

We record the engine binary by hash with each measurement. That is what let us
notice, in \S\ref{sec:tokens}, that a newer build delivers a quarter of the
expanded context of the build these scores were measured on.

The LongMemEval-S numbers come from an earlier campaign than the BEAM-1M ones.
We release its per-question records too, all eight runs of Table~\ref{tab:lme},
each carrying our answer, the judge's verdict, the passes used, the memories
delivered, and whether the supporting evidence was among them, keyed to the
benchmark's own question identifier. The last of those is the closest thing in this paper to a score for
retrieval on its own, and it is what \S\ref{sec:lme} uses to separate the engine
from the reader. We did not re-run these experiments for this paper. The corpus
is a public dataset and the harness grades it natively, so a reader
who wants an independent number can ingest it and run the released harness
without relying on ours.

\FloatBarrier
\section{Related work}

\paragraph{Why a memory system rather than a longer prompt.}
Retrieving evidence and giving it to a generator is the arrangement introduced
by retrieval-augmented generation \citep{lewis2020rag}, and the case for keeping
that arrangement as context windows grew has been made empirically. A model given
a long context does not use all of it equally, with material in the middle used
least reliably \citep{liu2024lost}, and the length a model handles in a synthetic
retrieval probe is consistently shorter than its advertised window
\citep{hsieh2024ruler}. That is the premise this paper works from, and it is also
BEAM's stated motivation.

\paragraph{Memory systems and the benchmarks built for them.}
Systems that manage what a model remembers across sessions include MemGPT, which
pages material between a small working context and external storage
\citep{packer2023memgpt}, MemoryBank, which adds a forgetting schedule
\citep{zhong2024memorybank}, and Mem0, which extracts and consolidates facts as a
conversation proceeds \citep{chhikara2025mem0}. \sys{} is narrower than all
three. It stores and returns, and does not decide what a memory means, which is
why every score in this paper is a score for a pair and we say so each time.
The benchmark side runs from multi-session open-domain dialogue
\citep{xu2022goldfish} through LoCoMo \citep{maharana2024locomo} and
LongMemEval \citep{wu2024longmemeval} to BEAM \citep{tavakoli2025beam}, which
supplies 100 conversations and 2{,}000 validated questions across ten ability
types at scales up to 10M tokens. Our runs use its 1M-token tier and its scoring
protocol, including its judge prompt and its rank correlation treatment of event
ordering.

Vendor reports on the same benchmark are the basis of Table~\ref{tab:compare}:
Exabase \citep{exabase2026beam}, Hindsight \citep{hindsight2026beam} and Mem0
\citep{mem02026state}. We take their figures as published and do not restate
them as controlled comparisons, for the reasons in \S\ref{sec:beam}.

\paragraph{Retrieval with and without term matching.}
BM25 \citep{robertson2009bm25} is the lexical baseline we run as a control
rather than as a component; the engine contains no lexical scoring. The dense
alternative has a long line behind it, from supervised passage retrievers
\citep{karpukhin2020dpr} to unsupervised ones \citep{izacard2022contriever},
with BEIR \citep{thakur2021beir} and MTEB \citep{muennighoff2023mteb} supplying
the evaluation suites that made the comparison routine. Section
\ref{sec:densebaseline} is an argument with a habit that line encourages, which
is to treat dense and language-independent as the same property.

\paragraph{Retrieval in many languages.}
Mr.\ TyDi \citep{zhang2021mrtydi} and MIRACL \citep{zhang2023miracl} established
multilingual passage retrieval as something measured across many languages at
once rather than English plus a sample, and NoMIRACL
\citep{thakur2024nomiracl} extended the question to what a system does when
retrieval returns nothing relevant. Multilingual encoders in wide use publish
their coverage as a list of languages \citep{feng2022labse, wang2024me5,
chen2024m3}, which is the fact \S\ref{sec:densebaseline} turns into a
measurement. All of this is text against text. Where a multilingual set is
obtained by translating an English one \citep{bonifacio2021mmarco}, the queries
carry the original's phrasing with them.

\paragraph{Finding a picture from a sentence.}
Retrieving a photograph from a sentence became a general capability with
contrastive image-text pretraining \citep{radford2021clip}, later refined at the
loss \citep{zhai2023siglip}. Our two open controls pair that
model's image encoder with its own English text encoder and with a multilingual
one distilled from it \citep{reimers2020multilingual}; the shared image side is
what lets us attribute the difference between the rows to the text side alone.
Crossmodal-3600 \citep{thapliyal2022xm3600} supplies the real photographs and
their captions in 36 languages, written independently by native speakers rather
than translated, which is what makes it usable for the question we ask of it.

\paragraph{Searching more than once, and grading the result.}
The re-ask protocol of \S\ref{sec:reask} sits beside iterative retrieval methods
\citep{xiong2021mdr, press2023compositionality, trivedi2023ircot} and differs
from them in refusing to put a model in the loop. Grading is by language model
in both benchmarks, following the paradigm surveyed by \citet{zheng2023judge}
and subject to the sensitivities documented since \citep{wang2024unfair}.
BEAM's per-question rubrics are nuggets in the retrieval-evaluation sense, and
automating their extraction and judging is an active line of its own
\citep{pradeep2025nuggets}.

\FloatBarrier
\section{Conclusion}

We set out to measure a production memory engine on two things that are usually
measured separately or not at all: how well it does on a large situational
retrieval benchmark, and whether its retrieval is genuinely independent of
language and of text.

On the first, \sys{} scores \pct{67.5} on BEAM-1M and \pct{95.7} on
LongMemEval-S, with 95\% question-sampling intervals of $[64.8, 70.2]$ and
$[93.4, 97.1]$. We place the BEAM figure beside other
published numbers and argue, with our own measurements, that the placement
should not be read as a ranking: the reader model is worth 2.0 points, the
re-ask budget up to 8.9, and neither is stated in the reports being compared.
The one controlled comparison we have is against our Scroll-tier engine read by
the same model, where the engine change is worth 1.4 points.

On the second, the useful result is not that we score highly but that the
comparison is bounded on both sides. Above, an open system built for this task
is not level across languages merely by being dense: one reaches \pct{91.0} on
English and \pct{4.7} on Russian, and the multilingual variant is level over the
languages it was trained on and collapses outside them. Below, the floor is
simpler. A photograph stored with no caption has no text to match, so a lexical method has nothing to score, and across 14 languages BM25 is exactly zero there while \sys{} reaches \pct{91.4}
recall@5. Where BM25 does have text to work with it reaches \pct{100} in the
caption's own language and at most \pct{9.7} in any other. Lexical retrieval can
be made multilingual by writing every caption in every language in advance,
which we measured too; it stops working at the first language nobody bought.

Three results run against us and we think they are the more useful half of the
paper. Low-resource languages degrade sharply: Swahili at \pct{53.0} and Telugu
at \pct{64.0} are our worst numbers anywhere. The open baselines reach \pct{6.0} and \pct{5.0} on
the same two, which says the difficulty is the field's and not ours alone, and
does not make our own numbers good. Attaching captions to photographs makes
cross-lingual retrieval worse by 11.4 points on average, and that one is a
defect in a shipping product and not an artifact of the benchmark: customer
stores hold exactly the mixture that produces it, and a photograph a user could
have found is being pushed out of the results by its neighbours' captions. We
have not fixed it. And one setting
omitted on the way into one stage cost 37 points of Korean top-1 accuracy on
short queries; we had called that stage's own limit and were wrong.

That last one produced the procedure we would most like to pass on. When one
language, or one condition, is much worse than the rest, measure the stage that
decides it on its own, with everything around it taken out of the path. If the
number matches the full system's, the limit is in that stage and you can stop
looking. If it does not, the limit is in the handling and it is findable. We ran
that test twice here and it returned a different verdict each time.

What we would pass on is smaller than a result. Three of the numbers we had been
publishing turned out to be wrong when we went back to the raw records for this
paper: a latency that averaged in a file our own script had marked contaminated,
a delivered-token figure that had never been counted, and an error count left
out of a denominator. None of them were caught by rerunning the analysis, because
the analysis was where they came from. They were caught by measuring the same
thing a second way and checking the two agreed. Where we did that, the paper is
worth something; where we did not, it is a report of what our own summaries said.

\clearpage
\addcontentsline{toc}{section}{References}
\bibliographystyle{plainnat}
\bibliography{refs}

\appendix

\FloatBarrier
\section{Prompts}
\label{app:prompts}

Reproducing a score requires the exact prompt, so these are given verbatim from
the source that produced the runs and are not paraphrased. The reader and judge
prompts for LongMemEval-S are also published on our model pages.

\subsection{BEAM-1M reader prompt}

Used for every BEAM-1M run in this paper.

\begin{verbatim}
Answer the question using only the memories below.
Do not use outside knowledge. If the memories do not contain the answer, say so.

MEMORIES
{memories}

QUESTION
{question}

Answer concisely.
\end{verbatim}

\subsection{BEAM-1M judge prompt}

The benchmark's own \texttt{unified\_llm\_judge\_base\_prompt}, used verbatim
from its repository with no modification, at temperature 0. We reproduce it by
reference rather than by copying, since the authoritative version is theirs and
a transcription here could drift from it.

\subsection{LongMemEval-S reader prompt}

\begin{verbatim}
Answer the question using ONLY the retrieved memories below (each is
prefixed with its [date]). This question is being asked on: {qdate}.
Apply whichever of these fits the question:
- For any 'how long ago' / 'how many days/weeks/months since' question,
  compute the duration relative to the asking date above (not any other
  today), using the memory dates.
- If the memories give CONFLICTING values for the same fact (different
  values as of different dates), mention BOTH and note which is more recent.
- If the question asks for ADVICE or a RECOMMENDATION, first identify this
  user's relevant preferences, interests, and past choices from the
  memories, then tailor your answer to them (not generic advice).
- Otherwise, answer the factual question concisely and directly.
If the answer is not in the memories, say you don't know. Answer in the
SAME LANGUAGE as the question.

Memories:
{mems}

Question: {q}

Answer:
\end{verbatim}

\noindent
This prompt is longer than the BEAM-1M one because it handles cases the
LongMemEval categories require: relative-time arithmetic against the asking
date, and conflicting values recorded at different dates. Both benchmarks were
run with the prompt shown for them and neither prompt was tuned against its
benchmark's rubric.

\subsection{LongMemEval-S judge prompt}

Run as a separate process from answer generation, at temperature 0, returning
only yes or no. \texttt{\{RULE\}} is the benchmark's own per-category rule,
applied unchanged.

\begin{verbatim}
I will give you a question, the correct answer, and a model's response.
{RULE} Respond with ONLY 'yes' or 'no'.

Question: {q}
Correct answer: {gt}
Model response: {ans}

Is the model response correct?
\end{verbatim}

\noindent
A verdict that is not an unambiguous yes or no leaves the question ungraded and
out of the denominator rather than being resolved by substring matching. That
choice matters in one direction: reading ``the answer is not yes'' as a yes
would raise every score it touched.

\end{document}